\documentclass[iop,onecolumn]{emulateapj}
\shortauthors{LAPI, SALUCCI, \& DANESE}
\shorttitle{STATISTICS OF DM HALOS}

\begin{document}

\title{Statistics of Dark Matter Halos from the Excursion Set Approach}
\author{A. Lapi$^{1,2}$, P. Salucci$^{2}$, L. Danese$^{2}$}
\affil{$^1$Dip. Fisica, Univ. `Tor Vergata', Via Ricerca Scientifica
1, 00133 Roma, Italy.}\affil{$^2$SISSA, Via Bonomea 265, 34136 Trieste,
Italy.}

\begin{abstract}
We exploit the excursion set approach in integral formulation to derive
novel, accurate analytic approximations of the unconditional and
conditional first crossing distributions, for random walks with
uncorrelated steps and general shapes of the moving barrier; we find the
corresponding approximations of the unconditional and conditional halo mass
functions for Cold Dark Matter power spectra to represent very well the
outcomes of state-of-the-art cosmological $N$-body simulations. In
addition, we apply these results to derive, and confront with simulations,
other quantities of interest in halo statistics, including the rates of
halo formation and creation, the average halo growth history, and the halo
bias. Finally, we discuss how our approach and main results change when
considering random walks with correlated instead of uncorrelated steps, and
Warm instead of Cold Dark Matter power spectra.
\end{abstract}


\section{Introduction}

\setcounter{footnote}{0}

In the standard cosmological framework, the seeds of cosmic structures like
quasars, galaxies and galaxy systems are constituted by dark matter (DM)
perturbations of the cosmic density field, originated by quantum effects
during the early inflationary Universe. The perturbations are amplified by
gravitational instabilities and, as the local gravity prevails over the
cosmic expansion, are enforced to collapse and virialize into bound `halos'.
In turn, these tend to grow hierarchically in mass and sequentially in time,
with small clumps forming first and then stochastically merging together into
larger and more massive objects. The halos provide the gravitational
potential wells where baryonic matter can settle in virial equilibrium, and
via a number of complex astrophysical processes (cooling, star formation,
feedback, etc.) originates the luminous structures that populate the visible
Universe.

To accurately describe and deeply understand the statistics of DM halos
constitute fundamental steps toward formulating a sensible theory of galaxy
formation and evolution. A milestone is constituted by the pioneering work of
Press \& Schechter (1974), who first provided an analytic expression for the
halo mass function, i.e., the halo abundance as a function of mass and
redshift; their theory envisages that a halo can collapse if it resided
within a sufficiently overdense region of an initial, Gaussian perturbation
field. However, the overdensity around a given spatial location depends on
the scale under consideration; thus the halo abundance can be computed from
the mass fraction in the density field which is above a critical threshold
for collapse when conveniently smoothed on different scales.

However, a relevant issue is to consider only those perturbations that
overcome the threshold on a given smoothing scale, but not on a larger one;
in fact, overdense regions embedded within a larger collapsing perturbation
are squeezed to smaller and smaller sizes, and care must be payed not to
double-count them. To cure this `cloud-in-cloud' problem, Bond et al. (1991)
developed the Excursion Set approach. This recognizes that the overdensity
around a given spatial location executes a random walk when considered as a
function of the smoothing scale (see Fig.~1); moreover, if the smoothing is
performed with a sharp filter in Fourier space, the steps of the random walk
are uncorrelated, i.e., the walk is Markovian. In this framework the
threshold for collapse plays the role of a barrier the random walk can hit;
the distribution of first crossing, i.e., the probability that a walk crosses
the barrier for the first time on a specific scale, can be closely related to
the halo mass function, and allows to avoid any double-counting.

The excursion sets approach has soon been exploited to derive the
`conditional' halo mass function (see Lacey \& Cole 1993), describing the
mass and redshift distribution of progenitors given they belong to a
descendant halo at a later time. This statistics has then been used to
generate merger trees, i.e., Montecarlo realizations of the halo merging
process (see Kauffmann \& White 1993; Somerville \& Kolatt 1999; Cole et al.
2000; Zhang et al. 2008a). With some additional assumptions, the excursion
set has been also helpful in developing models for the halo spatial
clustering and bias (see Mo \& White 1996; Sheth \& Lemson 1999). Subsequent
developments focused on the shape of the moving barrier; originally, a simple
model of halo spherical collapse (Gunn \& Gott 1972) was adopted, yielding a
constant barrier, independent of halo size. However, Gaussian perturbations
naturally possess a triaxial nature (Doroshkevich 1970), and the halo
collapse is likely to be ellipsoidal rather than spherical; as a result, the
threshold for collapse becomes a `moving', mass dependent barrier (Sheth et
al. 2001, Sheth \& Tormen 2002). The excursion set approach with moving
barriers has proven to be rather successful in reproducing reasonably well
the outcomes of cosmological $N$-body simulations (Sheth \& Tormen 1999,
Jenkins et al. 2001, Springel et al. 2005, Warren et al. 2006, Reed et al.
2007, Tinker et al. 2008; also Crocce et al. 2010, Angulo et al. 2012, Watson
et al. 2013). More recent developments concerns extension of the excursion
set approach to non-Markovian walks with correlated steps (Maggiore \& Riotto
2010a; Paranjape et al. 2012; Musso \& Sheth 2012), stochastic barriers
(Maggiore \& Riotto 2010b; Corasaniti \& Achitouv 2011), non Gaussian initial
conditions (Maggiore \& Riotto 2010c; de Simone et al. 2011a,b; Musso \&
Paranjape 2012), and application to the distribution of voids (Sheth \& van
de Weygaert 2004).

Despite these efforts, exact solutions to the first crossing distribution are
known only for simple barriers, like the constant or linear ones, but not for
the nonlinear shapes describing a realistic, ellipsoidal halo collapse. In
such cases, empirical formulae have been proposed to \emph{fit} the outcomes
of cosmological $N$-body simulations (see Sheth \& Tormen 2002; Parkinson et
al. 2008; Neistein \& Dekel 2008; Lam \& Sheth 2009; de Simone et al. 2011a).
Although such fitting formulae are inspired by the known solutions, being not
derived directly from the excursion set formalism may suffer of various
problems: inconsistencies with the excursion set approach in given mass and
redshift ranges; difficulties in the physical interpretation of the results;
dependence on the simulation setup; poor flexibility in describing comparably
well different barrier shapes; complex expressions, that make difficult to
derive analytically other statistical quantities of interest, like halo
merger rates and bias. One of the main goal of this paper is to provide
analytic approximations of the first crossing distribution (both
unconditional and conditional) for quite general shapes of the barrier, that
are derived directly from, and so are fully consistent with, the excursion
set formalism.

A related important issue we address concerns the halo creation rate, i.e.,
the change in abundance per unit time of newly created halos as a function of
mass and redshift. The seemingly simple task of deriving this quantity hides
delicate conceptual problems, and is to some extent still unresolved.
Naively, one can think of just taking the derivative of the halo mass
function; this is plainly not correct, because such a derivative is actually
the balance between the rate of halo creation by the merging of smaller ones,
and the rate of halo destruction by merging into larger ones (as stressed and
discussed by Cavaliere et al. 1991; see also Blain \& Longair 1993). Indeed,
the total derivative can be positive or negative, as a consequence of the
fact that small masses are dominated by the destruction and high masses by
the creation term. Some authors considered the possibility of taking the
`positive' term in the cosmic time derivative of the Press \& Schechter mass
function (e.g., Haehnelt et al. 1998); a partial justification for such an
approach was provided by Sasaki (1994), who obtained the same outcome by
requiring the destruction term to be scale-invariant. As a matter of fact,
$N$-body simulations show this assumption not to hold in general, and to
break down especially at small masses; in addition, Sasaki's procedure does
not provide consistent result for nonlinear barriers (see Mitra et al. 2011).

Another approach has been proposed by Sheth \& Pitman (1997); they described
the merging between halos in terms of the classic Smoluchowski's (1916)
binary coagulation equation (see also Silk \& White 1978), whose structure
naturally separates between the creation and destruction terms. In fact, as
already noted by Epstein (1983), there exists a strict analogy between the
excursion set approach with a white-noise power spectrum and a discrete
Poisson model of binary coagulation with additive kernel. However, in the
continuum limit relevant to the cosmological context, the creation and
destruction terms diverge, and replacing them with the finite, discrete
Poisson expressions lacks a mathematical justification. In addition, when
more realistic spectra are considered, like power-law or Cold DM ones, the
merger kernel from the excursion set is no longer symmetric, and the average
number of progenitors per merger event is greater than two; thus the analogy
with the binary coagulation equation does no longer hold (see Benson et al.
2005; Benson 2008; Neistein \& Dekel 2008; Zhang et al. 2008a).

Lacey \& Cole (1993) and Kitayama \& Suto (1996) attempted to derive an
expression for the creation rate directly from the excursion set framework;
taken at face value the result is divergent and calls for some form of
regularization. They sidestepped the problem by introducing the concept of
`formation' or major merger rate, according to which a halo is formed during
a merger if it acquires more than half its original mass. Percival \& Miller
(1999), basing on Bayes' theorem and assuming a flat prior on the
distribution of the threshold for collapse at fixed smoothing scale, obtained
an expression for the creation time distribution, which represents the
creation rate normalized in time (the normalization being again formally
infinite). All in all, these analysis failed to derive a sensible, finite
results for the creation rates; even the expressions provided for formation
rates and creation time distribution hold only for constant barriers, and
their extension to general moving barriers is nontrivial (see Mahmood \&
Rajesh 2005; also Moreno et al. 2009 for discussion). Another goal of this
paper is to provide finite, analytic expressions for the rates of halo
creation and formation with general moving barriers.

The plan of the paper is straightforward: in \S~2 we recall the basics of the
excursion set formalism; in \S~3 and 4 we derive our main results, i.e.,
approximated expression of the unconditional and conditional first crossing
distributions (and related mass functions) for general moving barriers; in
\S~5 we apply our main results to derive other quantities of interest in halo
statistics, including the rates of halo formation and creation, the average
halo growth history, and the halo bias for general moving barriers; in \S~6
we summarize our main findings. Appendix A provides a brief primer on the
$\zeta$-regularization technique for divergent integrals, while in Appendix B
and C we discuss how the excursion set approach and our main results change
when considering random walks with correlated instead of uncorrelated steps,
and with Warm instead of Cold Dark Matter power spectra.

Throughout this work we adopt a standard, flat cosmology (see Komatsu et al.
2011; Hinshaw et al. 2013; \textsl{Planck} Collaboration 2013) with matter
density parameter $\Omega_M = 0.3$, baryon density parameter
$\Omega_b=0.046$, and Hubble constant $H_0 = 100\, h$ km s$^{-1}$ Mpc$^{-1}$
with $h=0.70$.

\section{The excursion set approach}

In this Section we recall the basics of the excursion set approach, highlight
some delicate points useful in the sequel, and set the notation. The expert
reader may jump directly to \S~3.

We start by considering a spatial location in the Universe with comoving
coordinate $\mathbf{x}$ and local density contrast $\delta(\mathbf{x})\equiv
\rho(\mathbf{x})/\bar\rho-1$ over the background value $\bar\rho$. The
density contrast smoothed on a scale $R$ is given by
\begin{equation}
\delta(\mathbf{x},R)=\int{{\rm d}^3\mathbf{x'}}~W_R(|\mathbf{x}-\mathbf{x'}|)\,
\delta(\mathbf{x'})~,
\end{equation}
in terms of some filter function $W_R(\cdot)$. Since the behavior of
$\delta(\mathbf{x},R)$ as a function of $R$ is the interesting quantity,
without loss of generality we can pose $\mathbf{x}=0$ and denote
$\delta(0,R)$ just as $\delta(R)$. Passing to Fourier space and indicating
with a hat Fourier-transformed quantities, we can write
\begin{equation}
\delta(R)=\int{{\rm d}^3\mathbf{k}\over (2\pi)^3}~\hat W_R(k)\,
\hat\delta(\mathbf{k})~.
\end{equation}

We consider a Gaussian fluctuation field, whose statistical properties can be
completely encased into its power spectrum $P(k)$ defined as
\begin{equation}
\langle\hat \delta(\mathbf{k})\hat\delta(\mathbf{k'})\rangle = (2\pi)^3\,
P(k)\, \delta_D(\mathbf{k}+\mathbf{k'})~,
\end{equation}
with $\delta_D(\cdot)$ being the Dirac delta function, and
$\langle\cdot\rangle$ indicating the statistical average over the stochastic
variables $\hat\delta(\mathbf{k})$. In terms of the power spectrum, the
variance of the smoothed density field simply writes
\begin{equation}
S(R)=\int{{\rm d}^3\mathbf{k}\over (2\pi)^3}~ \hat W_R^2(k)\, P(k)~.
\end{equation}
For $P(k)$ we adopt the standard Cold DM shape by Bardeen et al. (1986; see
their Eq.~G3) with the correction for baryons by Sugiyama (1995; see his
Eq.~3.9), normalized in such a way that the r.m.s. $\sigma(R)\equiv
\sqrt{S(R)}$ takes on the value $\sigma_8=0.81$ on a scale $R=8\, h^{-1}$ Mpc
(consistent with the results by WMAP7/WMAP9 from Hinshaw et al. 2013 and by
the \textsl{Planck} Collaboration 2013).

As to the filter function, a common and convenient choice is a simple sharp
filter in Fourier space $\hat W_R(k)=\theta_{\rm H}(k_S-k)$, where
$\theta_{\rm H}(\cdot)$ is the Heaviside step function and $k_S(R)$ is a
cutoff wavenumber, to be related to the scale $R$. In such a case, indicating
with $\hat\delta(k)\equiv\int{\rm d}\Omega~\hat\delta(\mathbf{k})/4\pi$ the
polar-average in momentum space of $\hat\delta(\mathbf{k})$, one has
$\delta(R)=\int_0^{k_S(R)}{\rm d}k~k^2\, \hat\delta(k)/2\pi^2$ and
$S(R)=\int_0^{k_S(R)}{\rm d}k~k^2\, P(k)/2\pi^2$, hence ${\rm
d}_{k_S}\delta=k_S^2\, \hat\delta(k_S)/2\pi^2$ and ${\rm  d}_{k_S}S=k_S^2\,
P(k_S)/2\pi^2$ hold. Using these relations ${\rm d}_S\delta={\rm
d}_{k_S}\delta/{\rm  d}_{k_S}S=\hat\delta(k_S)/P(k_S)$ follows; moreover,
polar-averaging Eq.~(3) yields
$\langle\hat\delta(k_S)\,\hat\delta(k_{S'})\rangle=(2\pi)^3\, P(k_{S'})\,
\delta_D(k_S-k_{S'})/4\pi\,k_S^2=2\pi^2\, P(k_{S'})\,|{\rm  d}_{k_S}S|
\,\delta_D(S-S')/k_S^2= P(k_S)\,P(k_{S'})\, \delta_D(S-S')$.

All in all, $\delta(R)$ is seen to satisfies a Langevin equation with a Dirac
delta noise, in the form
\begin{equation}
{{\rm  d}\delta\over{\rm  d} S}={\hat\delta(k_S)\over P(k_S)}
~~~\mathrm{where}~~~~\langle{\hat\delta(k_S)\over
P(k_S)}{\hat\delta(k_{S'})\over P(k_{S'})}\rangle=\delta_D(S-S')~;
\end{equation}
this means that $\delta(R)$ executes a Markovian random walk as a function of
the smoothing scale $R$, with $S(R)$ playing the role of a time variable. In
more detail, the walk starts at $\delta=0$ when $S=0$ corresponding to large
values of $R$, and then as $R$ decreases and the pseudo-time $S(R)$
increases, $\delta(R)$ performs a stochastic motion under the influence of a
Gaussian noise with variance $S(R)$. An example is illustrated in Fig.~1.

A delicate point with the sharp filter in Fourier space is to associate a
mass $M$ to the smoothing scale $R$ (see also discussion in Appendix B). The
standard procedure, admittedly ambiguous to some extent, is to normalize the
expression of the filter in real space $W_R(r) = [\sin k_S r-k_S r\cos k_S\,
r]/2\pi^2 r^3$ to its maximum value $k_S^3/6\pi^2$, then to compute the
volume $V$ enclosed in the filter, and finally to set $M=\bar \rho\, V$. This
yields
\begin{equation}
M=4\pi\, \bar\rho\, \int_0^\infty{\rm d}r~r^2\, {6\pi^2\over k_S^3}\,
W_R(r)={12\pi\bar\rho\over k_S^3}\,\int_0^\infty{\rm d}x~\left[{\sin x\over
x}-\cos{x}\right]={6\pi^2\bar\rho\over k_S^3}~;
\end{equation}
In deriving the final expression we have used that $\int_0^\infty{\rm
d}x~\sin x/x=\pi/2$, and we have computed the Borel-regularized value of the
otherwise ill-defined, oscillating integral $\int_0^\infty{\rm d}x~\cos x$;
this amounts to introduce a regulator $\Lambda$ and to compute, in place of
the original integral, the well-defined quantity (see Hardy 1949 for details)
\begin{equation}
\lim_{\Lambda\rightarrow 0^+}\int_0^{\infty}{\rm d}x~e^{-\Lambda x}\,
\cos x=\lim_{\Lambda\rightarrow 0^+}\left[e^{-\Lambda
x}{\sin x-\Lambda\cos x\over 1+\Lambda^2}\right]_0^\infty=
\lim_{\Lambda\rightarrow 0^+}{\Lambda\over 1+\Lambda^2}=0~.
\end{equation}
All in all, posing $M=4\pi\,\bar\rho R^3/3$ one obtains the relation
$R=(9\pi/2)^{1/3}/k_S$ between the smoothing scale and the cutoff wavenumber.

In the excursion set formulation a halo is formed when the walk first crosses
a barrier $B(S,t)$ with general shape
\begin{equation}
B(S,t)= B_0+B_\gamma\, S^\gamma = \sqrt{q}\delta_c(t)\, \left\{1+\beta\,
\left[{q\delta_c^2(t)\over S}\right]^{-\gamma}\right\} =
\sqrt{S}\,(q\nu)^{1/2}\,\left[1+{\beta\over (q\nu)^{\gamma}}\right]~,
\end{equation}
with $B_0\equiv q^{1/2}\,\delta_c(t)$ and $B_\gamma\equiv \beta\,
q^{-\gamma+1/2}\,\delta_c(t)^{1-2\gamma}$. Here $\delta_c(t)$ is the critical
threshold for collapse extrapolated from linear perturbation theory. At the
current epoch $\delta_c\approx 1.686$ holds, although the precise value
weakly depends on cosmological parameters (see Eke et al. 1996); then it
evolves like $\delta_c(t)\propto 1/D(t)$ with the cosmological time $t$, in
terms of the linear growth function $D(t)$, see Weinberg (2008) for detailed
expressions.

In addition, the dependence of the barrier on the scale $S$ is specified by
the parameters $(q,\beta,\gamma)$, that are commonly set basing on theory of
halo collapse or by comparison with $N$-body simulations. The triple of
values $(1,0,0)$ corresponds to a constant barrier that describes spherical
collapse (Press \& Schechter 1974; Bond et al. 1991); the triple
$(0.707,0.47,0.615)$ corresponds to a nonlinear barrier that represents
ellipsoidal collapse (actually a collapse with the most probable values of
ellipticity and prolateness, see Sheth \& Tormen 2002 for details); finally,
the triple $(0.55,0.5,0.5)$ corresponds to a square-root barrier often
considered in the literature as a simpler description of ellipsoidal collapse
(Mahmood \& Rajesh 2005). In fact, note that for $\gamma>1/2$ not all walks
are guaranteed to cross the barrier (the r.m.s. $\delta$ of the walk scales
as $\sqrt{S}$), and barriers corresponding to different times can intersect
(an occurrence thought to represent fragmentation); thus one often adopt the
square-root barrier in place of the Sheth \& Tormen one, given that the two
are known to reproduce comparably well the outcomes of cosmological $N$-body
simulations.

The last equality in Eq.~(8) highlights that these barriers can be recast in
self-similar terms with the use of the variable $\nu\equiv \delta_c^2(t)/S$;
this is useful to treat at the same time the first crossing distribution for
different epochs and/or masses. In Fig.~2 we represent the shape $B(S)$ of
these barriers (color-coded) as a function of $\nu$. In the following we will
use these three barriers as prototypical examples to illustrate our results,
for simplicity referring to them as `constant', `square-root', and
`ellipsoidal'.

Then according to the excursion set prescription, the (unconditional) halo
mass function is given by
\begin{equation}
{{\rm d}N\over {\rm d}M} = {\bar \rho\over M^2}\, \left|{{\rm d log}S\over
{\rm d log}M}\right|\, S f(S)~;
\end{equation}
here $f(S)$ is the first crossing distribution, i.e., $f(S)\,
{\rm d}S$ represents the probability that a trajectory crosses the barrier
for the first time between $S$ and $S+{\rm d}S$. This quantity may be
determined as described next.

\subsection{The first crossing distribution}

It is convenient to formulate the problem in terms of the integral equation
(see Zhang \& Hui 2006)
\begin{equation}
\int_0^S{\rm d}S'~f(S')+\int_{-\infty}^{B(S)}{\rm d}\delta~P(\delta,S)=1~;
\end{equation}
here $B(S)$ is the moving barrier, $f(S)$ is the first crossing distribution,
and $P(\delta,S)$ is the probability for the trajectory to lie between
$\delta$ and $\delta+{\rm d}\delta$ at $S$. The above equation plainly states
that at a given $S$ a trajectory must either have crossed the barrier at some
smaller $S$ or still be below the barrier.

If no barrier were present, $P(\delta,S)$ would simply be a normal Gaussian
distribution with null mean and variance $S$, to read
\begin{equation}
P_0(\delta,S)={e^{-\delta^2/2S}\over \sqrt{2\pi S}}~;
\end{equation}
in presence of the barrier, to find $P(\delta,S)$ one has to subtract from
$P_0(\delta,S)$ the fraction of trajectories now at $(\delta, S)$ which have
crossed the barrier at some $S'<S$; this writes
\begin{equation}
P(\delta,S)=P_0(\delta,S)-\int_0^S{\rm d}S'~f(S')\,P_0[\delta-B(S'),S-S']~.
\end{equation}
Inserting Eqs.~(11) and (12) into Eq.~(10) and performing the integration
over $\delta$ yields
\begin{equation}
{\rm erfc}\left[{B(S)\over \sqrt{2 S}}\right]=\int_0^S{\rm d}S'~f(S')\,{\rm
erfc}\left[{B(S)-B(S')\over \sqrt{2 (S-S')}}\right]~,
\end{equation}
where ${\rm erfc}(\cdot)$ is the complementary error function. This
constitute a Volterra integral equation of the second kind for the unknown
function $f(S)$ given the barrier shape $B(S)$. Note that the integral
formulation of the excursion set approach expressed by Eq.~(13) holds for
Markovian random walks with uncorrelated steps; in Appendix B we will discuss
how the approach can be modified when some degree of correlation between the
steps is considered.

For a linear barrier $B_L(S)=B_0+B_1 S$ the solution of Eq.~(13) is easily
derived by applying a Laplace transformation, to obtain
\begin{equation}
\hat{f_L}(S)\equiv \int_0^\infty{\rm d}x~e^{-x\, S}\, f_L(x)=e^{-B_0\,
(B_1+\sqrt{B_1^2+2 S})}~;
\end{equation}
anti-transforming via the standard Bromwich integral (along a vertical
contour $\mathcal{C}$ in the complex plane having all the singularities of
$\hat{f_L}$ to the left; e.g., Arfken et al. 2013, p. 1038) leads to the
`inverse Gaussian' distribution
\begin{equation}
f_L(S)={1\over 2\pi i}\,\int_{\mathcal{C}}{\rm d}z~e^{z\,S}\,
\hat{f_L}(z)={B_0\over \sqrt{2\pi S^3}}\, e^{-B_L^2(S)/2 S}~;
\end{equation}
note that the solution for a constant barrier $B(S)=B_0$ is simply recovered
by placing $B_1=0$.

However, the barriers relevant to the collapse of DM halos are nonlinear,
featuring the typical shapes $B(S)=B_0+B_\gamma\, S^\gamma$ with
$1/2\leq\gamma\leq 1$ like in Eq.~(8); in these cases it is not possible to
solve exactly Eq.~(13) for the first crossing distribution, and one generally
recurs to numerical techniques (e.g., Zhang \& Hui 2006, Benson et al. 2013).
In fact, it is sufficient to discretize Eq.~(13) on a grid of values $S_i$
with spacing $\Delta S_i$ to obtain the solution by the recursive formula
\begin{equation}
f(S_i)={2\over \Delta S_{i-1}}\,\left\{{\rm erfc}\left[{B(S_i)\over \sqrt{2
S_i}}\right]-\sum_{j=0}^{i-1} f(S_j)\,{\rm erfc}\left[{B(S_i)-B(S_j)\over
\sqrt{2 (S_i-S_j)}}\right]\, {\Delta S_{j-1}+\Delta S_j\over 2}\right\}~,
\end{equation}
where $f(S_0)\equiv 0$ by definition.

\section{Unconditional first crossing distribution and halo mass function}

In this section we focus on the unconditional first crossing distribution. We
numerically solve Eq.~(16) and illustrate by the solid lines in Fig.~3 the
outcomes for constant, square-root and ellipsoidal barriers (color-coded). In
addition, the crosses show the outcome of state-of-the-art cosmological
$N$-body simulations (fit to Tinker et al. 2008; see \S~1 for other
references). This is in close agreement to both the square-root and the
ellipsoidal distribution, actually striking an intermediate course between
the two; the constant barrier result is instead substantially different.

Here our aim is to derive useful approximate expressions to the unconditional
first crossing distribution in the limit $S\ll \delta_c^2(t)$, i.e., large
masses and/or early times; in self-similar terms this corresponds to the
limit $\nu\gg 1$. We start by considering in Eq.~(13) the limit for small
$S$: on the lhs the quantity $B(S)/\sqrt{2 S}$ tends to infinity so that we
can exploit the asymptotic expansion of the function ${\rm erfc}(x)\simeq
e^{-x^2}/x\sqrt{\pi}$ for large argument; meanwhile inside the integral on
the rhs the quantity $[B(S)-B(S')]/\sqrt{2 (S-S')}\simeq \gamma B_\gamma
S^{\gamma-1/2} \sqrt{(S-S')/2 S}$ tends to zero for $\gamma\geq 1/2$ so that
we can expand the function ${\rm erfc}(x)\simeq 1-2 x/\sqrt{\pi}$ for small
values of its argument.

Differentiating both sides of Eq.~(13) with respect to $S$ yields, to the
lowest order,
\begin{equation}
{B_0\, e^{-B^2(S)/2 S}\over \sqrt{2\pi S^3}}\, \left[1+(1-2\gamma)\,
{B_\gamma\over B_0}\, S^\gamma\right]\simeq f(S)-{B_\gamma\, \gamma\,
S^{\gamma-1}\over \sqrt{2\pi}}\, \int_0^S{\rm d}S'\, {f(S')\over
\sqrt{S-S'}}\, \left[1+2\, (\gamma-1)\,{S-S'\over S}\right]~.
\end{equation}
The structure of this equation suggest to consider the following ansatz for
the small $S$ expansion of the first crossing distribution
\begin{equation}
f(S)\simeq {B_0\, e^{-B^2(S)/2 S}\over \sqrt{2\pi S^3}}\,
\left[1+k_\gamma\,{B_\gamma\over B_0}\,S^\gamma\right]~,
\end{equation}
where $k_\gamma$ is a constant, dependent on $\gamma$, to be determined.
Using it in Eq.~(17), and approximating the integral by the Laplace method
(major contribution is from a small interval around $S'\simeq S$), we obtain
to the lowest order
\begin{equation}
{B_0\, e^{-B^2(S)/2 S}\over \sqrt{2\pi S^3}}\, \left[1+(1-2\gamma)\,
{B_\gamma\over B_0}\, S^\gamma\right]\simeq {B_0\, e^{-B^2(S)/2 S}\over
\sqrt{2\pi S^3}}\, \left[1+(k_\gamma-\gamma)\,{B_\gamma\over
B_0}\,S^\gamma\right]~.
\end{equation}
This implies $k_\gamma=1-\gamma$ and thus the approximate first crossing
distribution writes
\begin{equation}
f(S)\simeq {B_0\, e^{-B^2(S)/2 S}\over \sqrt{2\pi S^3}}\,
\left[1+(1-\gamma)\,{B_\gamma\over B_0}\,S^\gamma\right]~;
\end{equation}
note that for the linear ($\gamma=1$) and constant ($B_\gamma=0$) barriers
one recovers the exact solution Eq.~(15).

Specializing to the barrier shape given by Eq.~(8) and passing to the
self-similar variable $\nu$ we have
\begin{equation}
f(\nu)\simeq \sqrt{q\over
2\pi\nu}\,e^{-q\nu\,[1+\beta\,(q\nu)^{-\gamma}]^2/2}\,
\left[1+{(1-\gamma)\,\beta\over (q\nu)^{\gamma}}\right]~;
\end{equation}
this approximate expression valid for $\nu\gg 1$ and for a generic $\gamma$
constitutes a novel result. In Fig.~3 we compare these approximations (dashed
lines) of the unconditional first crossing distribution to the exact results
(solid lines) for the constant, square-root, and ellipsoidal barriers.
Remarkably, the approximations perform very well over a wide range in $\nu$,
and specifically better than $\la 10\%$ for $\nu\ga 0.3$ (and better than
$\la 30\%$ for $\nu\ga 0.1$). In terms of the characteristic mass
$M_\star(z)$ defined by $\delta_c^2(z)=S[M_\star(z)]$, for a standard Cold DM
spectrum the range translates into $M\ga 10^{-2}\, M_\star$ at $z\approx 0$,
extends to $M\ga 10^{-3}\, M_\star$ at $z\approx 1$ and to $M\ga 3\times
10^{-4}\, M_\star$ at $z\approx 2$, practically encompassing all masses of
cosmological interest for $z\ga 2$.

In Fig.~4 we illustrate the corresponding mass function at different
redshifts (coded with linestyles), computed by inserting in Eq.~(9) our
approximated expressions Eq.~(20) for the first crossing distribution. When
compared to the outcomes of cosmological $N$-body simulations (crosses; fit
to Tinker et al. 2008) for Cold Dark Matter, the constant barrier mass
function considerably underestimate the abundance of simulated halos,
especially at high $z$; the square-root and above all the ellipsoidal barrier
mass functions perform instead substantially better.

We note that our approximation to the halo mass function based on Eq.~(20)
and (9) is extremely flexible with respect to changes of the power spectrum.
For example, in Appendix C we will exploit it for Warm Dark Matter, and show
its good agreement with recent $N$-body simulations.

\section{Conditional first crossing distribution and halo mass function}

In this section we focus on the conditional first crossing distribution,
i.e., the probability of an object to feature a mass in the range between
$M'$ and $M'+{\rm d}M'$ at time $t'$ provided that it becomes incorporated
into a larger object of mass $M>M'$ at a later time $t>t'$. As can be easily
understood on the basis of Fig.~1, in the excursion set approach this two
barrier problem is equivalent to determine the first crossing distribution of
a random walk with a single barrier of the form $B(\Delta
S,t,t')=B(S',t')-B(S,t)$ and pseudo-time variable $\Delta S=S'-S$. Note that
if the barrier $B(S,t)=\delta_c(t)$ were constant, then $B(\Delta
S)=\delta_c(t')-\delta_c(t)=\Delta\delta_c$ would apply, and the result would
obtain just rescaling the unconditional distribution from the self-similar
variable $\nu\equiv \delta_c^2/S$ to $\nu_c\equiv (\Delta\delta_c)^2/\Delta
S$.

In Fig.~5 we illustrate as solid lines the exact solutions by solving
numerically Eq.~(16), as a function of $\nu_c$, for a descendent mass $M=
10^{13}\, M_{\odot}$ and redshift interval $\Delta z=1$. In fact, note that
the conditional distributions, except for constant barrier case, do not
depend solely on $\nu_c$ but also slightly on the descendant halo mass and
redshift; the rescaled unconditional mass functions are plotted for reference
in Fig.~5 as dotted lines.

Here our aim is to derive useful approximated expressions for the conditional
first crossing distributions. Since $B(\Delta S,t,t')$ is a weakly nonlinear
barrier, we can approximate it as $B(\Delta S)\simeq C_0+C_1\, \Delta S+ C_2
(\Delta S)^2$, and the corresponding first crossing distribution as $f(\Delta
S)\simeq f_L(\Delta S)+C_2\,\partial_{C_2} f_{|C_2=0}$, where $f_L(\Delta S)$
is the inverse Gaussian distribution of Eq.~(15). The unknown quantity
$\partial_{C_2} f_{|C_2=0}$ may be derived along the following steps. First,
we apply the operator $\partial_{C_2} (\cdot)_{|C_2=0}$ to Eq.~(13) and
obtain
\begin{eqnarray}
\nonumber -\sqrt{2}\,(\Delta S)^{2}\,{e^{-(C_0+C_1 \Delta S)^2/2
\Delta S}\over \sqrt{\pi\Delta S}}=\int_0^{\Delta S}{\rm d}\Delta S'~
\partial_{C_2} f_{|C_2=0}(\Delta
S')\,{\rm erfc}\left({C_1\,\sqrt{\Delta S-\Delta S'\over 2}}\right)-\\
\\
\nonumber ~~~~~~~~~~-\sqrt{2\over\pi}\,\int_0^{\Delta S}{\rm d}\Delta
S'~f_L(\Delta
S')\, {(\Delta S)^2-(\Delta S')^2\over \sqrt{\Delta S-\Delta S'}}\,
e^{-C_1^2\, (\Delta S-\Delta S')/2}~;
\end{eqnarray}

Then we perform a Laplace transformation term by term. To this purpose, we
use the following Laplace rules (see, e.g., Abramowitz \& Stegun 1972, p.
1020): $\int_0^x{{\rm d}x'}~F(x-x')\, G(x') \Rightarrow \hat F(s)\, \hat
G(s)$, $e^{k\,x}\, F(x)\Rightarrow \hat F(s-k)$,
$(-1)^n\,x^n\,F(x)\Rightarrow {\rm d}^n\hat{F}(s)/{\rm d}s^n$, holding for
generic functions $F(x), G(x)$ and their transforms $\hat F(s), \hat G(s)$,
for any real number $k$ and for any integer $n$. We also use the following
Laplace transforms of elementary functions: $e^{-k^2/4 x}/\sqrt{\pi x}
\Rightarrow e^{-k\sqrt{s}}/\sqrt{s}$, $e^{k^2 x}\,{\rm
erfc}(k\sqrt{x})\Rightarrow 1/\sqrt{s}\,(\sqrt{s}+k)$, $k\, e^{-k^2/4
x}/2\sqrt{\pi x^3}\Rightarrow e^{-k \sqrt{s}}$, $2\,\sqrt{x/\pi}\Rightarrow
s^{-3/2}$, holding for any real number $k>0$.

After some tedious algebra, we get
\begin{eqnarray}
\nonumber -\sqrt{2}\, e^{-C_0\,C_1}\, {{\rm d}^2\over{\rm d}(\Delta
S')^2}\,\left[{e^{-C_0\, \sqrt{2\Delta S'}}\over \sqrt{\Delta
S'}}\right]_{|\Delta S' = \Delta S+C_1^2/2} = {\hat{\partial_{C_2}
f_{|C_2=0}}(\Delta S)\over [\Delta S'+C_1\,\sqrt{\Delta
S'/2}]_{|\Delta S' = \Delta S+C_1^2/2}}+\\
\\
\nonumber ~~~~~~~~~~+e^{-C_0\, C_1}\left\{{{\rm d}\over {\rm d}(\Delta
S')}\left[{e^{-C_0\sqrt{2\Delta S'}}\over \sqrt{2}\,(\Delta
S')^{3/2}}\right]-C_0\,{e^{-C_0\sqrt{2\Delta S'}}\over 2\,(\Delta S')^2}
\right\}_{|\Delta S' = \Delta S+C_1^2/2}~.
\end{eqnarray}
Rearranging the various terms yields
\begin{equation}
\hat{\partial_{C_2} f_{|C_2=0}}(\Delta S) = - C_0\, e^{-C_0\,
(C_1+\sqrt{C_1^2+2
\Delta S})}\, \left[{C_0\over (C_1^2+2 \Delta S)^{1/2}}+{1+C_0 C_1\over
C_1^2+2 \Delta S}+{C_1\over (C_1^2+2 \Delta S)^{3/2}}\right]~.
\end{equation}

Then we anti-transform term by term. To this purpose, we use the following
inverse Laplace transforms of elementary functions:
$e^{-k\sqrt{s}}/\sqrt{s}\Rightarrow e^{-k^2/4 x}/\sqrt{\pi x}$ ,
$e^{-k\sqrt{s}}/s \Rightarrow {\rm erfc}(k/2\sqrt{x})$,
$e^{-k\sqrt{s}}/s^{3/2}\Rightarrow 2\sqrt{x/\pi}\, e^{-k^2/4\, x}-k\,{\rm
erfc}(k/2\sqrt{x})$, holding for any real number $k>0$. This leads to
\begin{equation}
\partial_{C_2} f_{|C_2=0}(\Delta S) = - C_0 e^{-(C_0+C_1 \Delta S)^2/2 \Delta
S}\,
\left\{{e^{C_0^2/2 \Delta S}\over 2} {\rm erfc}\left(C_0\over \sqrt{2
\Delta S}\right)+{C_0+C_1\,\Delta S\over \sqrt{2\pi \Delta S}}\right\}~;
\end{equation}
finally, we find
\begin{equation}
f(\Delta S) \simeq {C_0 e^{-(C_0+C_1\Delta S)^2/2 \Delta S}\over\sqrt{2 \pi
\Delta S^3}}\, \left\{1-C_2\,(\Delta S)^{3/2}\, \left[\sqrt{\pi\over
2}\,e^{C_0^2/2 \Delta S} {\rm erfc}\left(C_0\over \sqrt{2 \Delta
S}\right)+{C_0+C_1 \Delta S\over \sqrt{\Delta S}}\right]\right\}~.
\end{equation}

We now specialize our computation to the popular family of barrier shapes of
Eq.~(8), in terms of the normalized coefficients
\begin{eqnarray}
&&\nonumber \tilde C_0\equiv {C_0\over S^{1/2}}=\left[q{\delta_c^2(t')\over
S}\right]^{1/2}-\left[q{\delta_c^2(t)\over
S}\right]^{1/2}+\beta\left\{\left[q{\delta_c^2(t')\over
S}\right]^{-\gamma+1/2}-\left[q{\delta_c^2(t)\over
S}\right]^{-\gamma+1/2}\right\}~,\\
&&\nonumber\\
&&\tilde C_1\equiv C_1\, S^{1/2}=\beta\gamma\,\left[{q\delta_c^2(t')\over
S}\right]^{-\gamma+1/2}~,\\
&&\nonumber\\
&&\nonumber \tilde C_2\equiv C_2\, S^{3/2}=-{\beta\gamma\,(1-\gamma)\over
2}\,\left[{q\delta_c^2(t')\over S}\right]^{-\gamma+1/2}~;
\end{eqnarray}
Using these expressions, the (approximate) conditional first crossing
distribution takes the form
\begin{eqnarray}
\nonumber f(\Delta S) &\simeq& {\tilde C_0 e^{-[\tilde C_0+\tilde C_1\Delta
S/S]^2/2
\Delta S/S}\over\Delta S\sqrt{2 \pi
\Delta S/S}}\, \left\{1-\tilde C_2\,\left({\Delta S\over
S}\right)^{3/2}\times\right.\\
&&\\
\nonumber &\times&\left. \left[\sqrt{\pi\over 2}\,e^{\tilde C_0^2/2 \Delta
S/S} {\rm erfc}\left(\tilde C_0\over \sqrt{2 \Delta S/S}\right)+\tilde C_0\,
\sqrt{S\over \Delta S}+\tilde C_1\sqrt{\Delta S\over S}\right]\right\}~.
\end{eqnarray}
This is a novel result, that generalize to finite time difference the
expression of Zhang et al. (2008b).

In Fig.~5 we compare this approximation (dashed lines) for the conditional
first crossing distribution to the exact results (solid line) for the
constant, square-root, and ellipsoidal barriers. Remarkably, the
approximation perform very well over a wide range in $\nu_c$, and
specifically better than $\la 10\%$ for $\nu_c\ga 0.3$; in terms of halo
masses, for a standard Cold DM spectrum the range translates into $M'/M\ga
10^{-3}$ with $\Delta z=1$. Note that the mass range where the approximation
works to this accuracy level becomes larger for higher descendant mass and
smaller redshift difference; actually, for $\Delta z\la 0.1$, the
approximation works very well for all the masses of cosmological interest.

Once the conditional first crossing distribution is known, the conditional
mass function can be computed as
\begin{equation}
{{\rm d}N\over {\rm d}M'}(M'\rightarrow M, \Delta t) = {M\over M'^2}\, \Delta
S f(\Delta S)\, \left|{{\rm d log}\Delta S\over {\rm d log}M'}\right|~.
\end{equation}
In Fig.~6 we illustrate the outcomes at different redshifts
(linestyle-coded), computed by inserting our approximation Eq.~(28) into
Eq.~(29). Note that for small redshift difference $\Delta z\la 0.1$ our
result concurs with that by Zhang et al. (2008b).

\section{Applications to halo statistics}

In this section we apply our main results given by Eq.~(20) and (28) to study
various quantities of interest in halo statistics, and compare them with the
outcomes of $N$-body simulations.

\subsection{Rates of Halo Creation}

The rates at which DM halos of given mass are created by merging of smaller
ones is an essential ingredient in galaxy formation and evolution models. To
recall some examples, these rates have been used to investigate AGN/quasar
activity (e.g., Wyithe \& Loeb 2003; Mahmood et al. 2005; Lapi et al. 2006),
FIR/submm-selected galaxies (e.g., Lapi et al. 2011; Cai et al. 2013), Lyman
break galaxies and cosmic reionization (e.g., Kolatt et al. 1999; Mao et al.
2007), abundances of binary supermassive black holes and related coalescence
rates (e.g., Milosavljevic \& Merritt 2001; Volonteri et al. 2002),
astrophysics of first stars (e.g., Santos et al. 2002; Scannapieco et al.
2003), galaxy clustering (e.g., Percival et al. 2003; Xia et al. 2012),
statistics of galaxy-galaxy gravitational lensing (e.g., Lapi et al. 2012).

As discussed in the references above (see also Cavaliere et al. 1991, Blain
\& Longair 1993, Sasaki 1994, Haehnelt et al. 1998), to address these issues
one cannot rely on the total derivative ${\rm d}N/{\rm d}t$ of the halo mass
function, because it includes a balance between the rate of halo creation by
merging of smaller ones, and the rate of halo destruction by merging into
larger ones. Indeed, the total derivative can be positive or negative, as a
consequence of the fact that small masses are dominated by the destruction
and high masses by the creation term. Extracting from ${\rm d}N/{\rm d}t$ the
rates of halo creation is actually a non-trivial task.

Such a delicate issue can be attacked with the excursion set approach (see
Lacey \& Cole 1993, Kitayama \& Suto 1996, Mahmood \& Rajesh 2005, Moreno et
al. 2009). The problem is that in such a framework all halos continuously
merge with other ones, and therefore are newly-born to some extent; as shown
below, pursuing this viewpoint leads to divergent expressions for the
creation rates, that need to be regularized. On the other hand, to obtain a
finite answer without the need for a regularization, some of the
aforementioned authors have envisaged that a new halo is born only when its
change in mass during a merger has been substantial, i.e., the halo has
undergone a major merger. In the above literature such major merger rates are
usually dubbed `formation rates', to distinguish them from the `creation
rates' that instead are computed irrespective of the mass change during the
merger event. Given that both these quantities are widely used in the
literature and extremely relevant for galaxy formation models, our purpose
here is to provide novel analytic expressions of the creation and formation
rates for general moving barriers.

We start by computing the rate ${\rm d}^2 p_{M'\rightarrow M}/ {\rm d}M' {\rm
d}t$ at which a halo of mass $M$ originates from a progenitor halo with mass
in the range between $M'$ and $M'+{\rm d}M'$ over an infinitesimal timestep
${\rm d}t$ (e.g., Kitayama \& Suto 1996, their Eq.~8); in other words, this
quantity represents the progenitor distribution for infinitesimal look-back
times. It can be derived from the above expression Eq.~(28) on taking the
limit $t'\rightarrow t$ and changing variable from $\Delta S$ to $M'$, i.e.,
${\rm d}^2 p_{M'\rightarrow M}/ {\rm d}M' {\rm d}t \equiv \lim_{t'\rightarrow
t}\, f(\Delta S)\, |{\rm d}\Delta S/{\rm d}M'|/\Delta t$. In such a limit
$\tilde C_0\simeq \bar C_0\, |\dot\delta_c|\, {\rm d} t/\sqrt{S}\rightarrow
0$, and the result reads\footnote{Note that the terms proportional to $\tilde
C_0\, \sqrt{S/\Delta S}\rightarrow \bar C_0\, |\dot\delta_c|\, {\rm
d}t/\sqrt{\Delta S}$ appearing for $t'\rightarrow t$ in the exponential and
in the square bracket of Eq.~(28) are subdominant since ${\rm d}t\rightarrow
0$, for any finite value of $\Delta S$. On the other hand, if $\Delta S$
would also tend to zero (corresponding to $M'\rightarrow M$), then the limit
would become ill-defined and some form of regularization would be needed. We
will come back to the issue below in this section when dealing with the
creation rates.}
\begin{equation}
{{\rm d}^2 p_{M'\rightarrow M}\over {\rm d}M' {\rm d}t} = {\bar C_0\,|\dot
\delta_c(t)| e^{-\bar C_1^2 \Delta S/2\, S}\over\sqrt{2 \pi (\Delta S)^3}}\,
\left|{{\rm d}S\over {\rm d}M}\right|_{M'}\,\left\{1-\bar C_2 \left({\Delta
S\over S}\right)^{3/2}\, \left[\sqrt{\pi\over 2}+ \bar C_1 \left({\Delta
S\over S}\right)^{1/2} \right]\right\}~,
\end{equation}
in terms of the quantities
\begin{eqnarray}
&&\nonumber \bar C_0\equiv \sqrt{q}\,
\left[1-\beta\,(2\gamma-1)\,(q\nu)^{-\gamma}\right]~,\\
&&\nonumber\\
&&\bar C_1\equiv \beta\,\gamma\,(q\nu)^{-\gamma+1/2}~,\\
&&\nonumber\\
&&\nonumber \bar C_2\equiv -{\beta\gamma\,(1-\gamma)\over
2}\,(q\nu)^{-\gamma+1/2}~
\end{eqnarray}
that depend only on the self-similar variable $\nu=\delta_c^2(t)/S$. The
above expression coincides with that obtained by Zhang et al. (2008b).

Then the halo formation rates are given by (see, e.g., Kitayama \& Suto 1996,
their Eqs.~8 and 13)
\begin{equation}
\partial_t N_+^{\rm form}=N\, \int_0^{M_f}{\rm d}M'~{{\rm d}^2 p_{M'\rightarrow
M}\over {\rm d}M' {\rm d}t}
\end{equation}
where $M_f$ is a suitable fraction of the final mass $M$, typically
$M_f\approx M/2$; as mentioned at the beginning of this Section, the
rationale of the above expression is that a new halo is formed if more than
half of its final mass is acquired during the merger, i.e., if the merger has
been a `major' one. Note that to simplify the notation we have indicated the
unconditional mass function with $N$ in place of ${\rm d}N/{\rm d}M$. Using
Eq.~(30) and changing variable from $M'$ to $x\equiv\sqrt{S/\Delta S}$ yields
\begin{equation}
\partial_t N_+^{\rm form}={2 \bar C_0 |\dot\delta_c|\, N\over \sqrt{2\pi S}}\,
\int_0^{1/\sqrt{S_{M_f}/S_M-1}}{\rm d}x~e^{-\bar C_1^2/2 x^2}\,
\left[1-\sqrt{\pi\over
2}\,{\bar C_2\over x^3}-{\bar C_1\bar C_2\over x^4}\right]~.
\end{equation}
Remarkably, the integrals can be carried out analytically and we obtain
\begin{eqnarray}
\nonumber\partial_t N_+^{\rm form}&=&{2 \bar C_0 |\dot\delta_c|\, N\over
\sqrt{2\pi S}}\,\left[e^{-\bar C_1^2\, s_f/2}\, \left({1\over
s_f^{1/2}}-\sqrt{\pi\over 2}\,{\bar C_2\over \bar
C_1^2}-{\bar C_2\over \bar C_1}\, s_f^{1/2}\right)\right.+\\
& &\\
\nonumber&-&\left.\sqrt{\pi\over 2}\,\left(\bar C_1+{\bar C_2\over \bar
C_1^2}\right)\, {\rm erfc}\left(\sqrt{\bar C_1^2\, s_f\over
2}\right)\right]~,
\end{eqnarray}
with $s_f\equiv S_{M_f}/S_M-1$. This expression for a general moving barrier
constitutes a novel result. In Fig.~7 we illustrate the formation rates as a
function of halo mass at different redshifts, for the constant, square-root
and ellipsoidal barriers.

On the other hand, the halo `creation' rates are given by (see, e.g.,
Kitayama \& Suto 1996, their Eqs.~8 and 10)
\begin{equation}
\partial_t N_+^{\rm crea}=N\,\int_0^{M}{\rm d}M'~{{\rm d}^2 p_{M'\rightarrow M}
\over {\rm d}M' {\rm d}t}~;
\end{equation}
this constitutes the limit for $M_f\rightarrow M$ in the expression for the
formation rates; as mentioned at the beginning of this Section, the rationale
of the above expression is that a new halo is continuously created by merging
of smaller ones, irrespective of the mass change during the merger. Plainly,
the integral in Eq.~(35) does not converge, because the term proportional to
$s_f^{-1/2}$ in Eq.~(34) blows up when $s_f\rightarrow 0$ (and since we are
considering infinitesimal time-steps ${\rm d}t$, even the expression of the
integrand in Eq.~35 would be ill-defined around $M'\rightarrow M$, cf. the
footnote concerning Eq.~30). However, this divergence (or ill-definiteness)
is easily recognized to be unphysical, being related to the continuum nature
of this treatment where infinitesimal mass changes can take place. In fact,
one is actually counting the transition of an object into itself; this has
been pointed out for a constant barrier by Kitayama \& Suto (1996), by
showing that the same divergence is present in the destruction rate (that can
be obtained from the creation rate and the unconditional distribution using
Bayes' theorem), and that the two exactly cancels out in the difference of
the rates to yield the total derivative of the mass function.

Here we derive a finite result for the creation rate on exploiting the
$\zeta-$regularization technique for divergent integrals; a brief primer on
the latter is provided in Appendix A. First of all, we introduce in Eq.~(35)
a cutoff $\Lambda$, and rewrite conveniently the upper integration limit as
$M\,(1+1/\Lambda^2)$. Then we change variable from $M'$ to
$x\equiv\sqrt{S/\Delta S}$; this turns the upper limit of integration into
$[S_{M(1+1/\Lambda^2)}/S_M-1]^{-1/2}\simeq \Lambda |{\rm d}\log M/{\rm d}\log
S|^{1/2}$, and we obtain
\begin{equation}
\partial_t N_+^{\rm crea}={2 \bar C_0 |\dot\delta_c|\, N\over \sqrt{2\pi S}}\,
\lim_{\Lambda\rightarrow\infty}
\int_0^{\Lambda\, |{\rm d}\log M/{\rm d}\log S|^{1/2}}{\rm d}x~e^{-\bar C_1^2/2
x^2}\,
\left[1-\sqrt{\pi\over 2}\,{\bar C_2\over x^3}-{\bar C_1\bar C_2\over
x^4}\right]~.
\end{equation}
Only the first integral diverges, and according to the procedure outlined in
Appendix A, we find the $\zeta$-regularized value
\begin{equation}
\lim_{\Lambda\rightarrow\infty} \int_0^{\Lambda\, |{\rm d}\log M/{\rm d}\log
S|^{1/2}}{\rm d}x~e^{-\bar C_1^2/2 x^2} \cong {1\over 2}\, \left|{{\rm d}\log
M\over {\rm d}\log S}\right|^{1/2}-\sqrt{\pi\over 2}\, \bar C_1.
\end{equation}
Finally, we obtain for the creation rates
\begin{equation}
\partial_t N_+^{\rm crea}={\bar C_0 |\dot\delta_c|\, N\over \sqrt{2\pi S}}\,
\left\{\left|{{\rm d}\log
M\over {\rm d}\log S}\right|^{1/2}-\sqrt{2\pi}\,\left(\bar C_1+2\, {\bar
C_2\over \bar C_1^2}\right)\right\}= {\sqrt{q}\,|\dot\delta_c|\, N\over
\sqrt{2\pi S}}\,g(\nu)~;
\end{equation}
here we have highlighted for compactness the factor $g(\nu)$, which depends
only on the self-similar variable $\nu=\delta_c^2/S$ at given power spectrum.
The expression of Eq.~(38) constitutes a novel result. In Fig.~8 we
illustrate the creation rates as a function of halo mass at different
redshifts, for the constant, square-root and ellipsoidal barriers.

Now to compare our results with $N$-body simulations (see Moreno et al.
2009), we compute the creation time distributions $c(t|m)$; this is defined
as the time normalized creation rate
\begin{equation}
c(t|m) = {\partial_t N_+^{\rm crea}\over \int_0^{\infty}{\rm d}t~\partial_t
N_+^{\rm crea}}~.
\end{equation}
Using our expressions Eq.~(21) and (38) we can write
\begin{equation}
c(\nu) = {c(t|m)\over |\dot \delta_c|}\, |\partial_\nu{\delta_c}| =
{\sqrt{\nu}\, f(\nu)\,g(\nu)\over \int_0^{\infty}{\rm d}\nu~\sqrt{\nu}\,
f(\nu)\,g(\nu)}~;
\end{equation}
this shows that the creation time distribution can be put in self-similar
form. In Fig.~9 we illustrate the outcomes for different barriers with the
solid lines. Note that our result for non-constant barriers is somewhat
different from that obtained by Percival \& Miller (1999) on using Bayes'
theorem and assuming a uniform $\delta_c$ distribution at given $S$; their
outcome is still given by Eq.~(40) but with $g(\nu)=1$, and is illustrated by
the dotted lines in Fig.~9; the crosses refer to the results of cosmological
$N$-body simulations (see Moreno et al. 2009). We find a very good agreement
between the $N$-body data and our result for the ellipsoidal barrier.

Finally, for the sake of completeness, we also compare the formation and
creation rates with two their proxies often adopted in the literature; it is
beyond the scope of the present paper to discuss their underlying theoretical
background (but see \S~1 for a quick overview). The first is obtained by
taking the `positive' term in the cosmic time derivatives of the mass
function (see Haehnelt et al. 1998, Blain \& Longair 1993, Sasaki 1994);
using our approximated expression in Eq.~(21) one obtains
\begin{equation}
\partial_t N_+^{\rm pos} = {|\dot\delta_c|\, N\over \delta_c}\,
\left\{q\nu\left[1+{\beta\over
(q\nu)^\gamma}\right]^2+{2\beta\gamma(1-\gamma)\over \beta
(1-\gamma)+(q\nu)^\gamma}\right\}~.
\end{equation}
For brevity we will refer to this prescription as `positive' rate. The second
has been proposed by Sheth \& Pitman (1997), based on the analogy between the
excursion set approach with a white-noise power spectrum and a discrete
Poisson model of binary coagulation with additive kernel. Their prescription
writes
\begin{equation}
\partial_t N_+^{\rm coag} = {|\dot\delta_c|\, N\over \delta_c}\nu ~
\end{equation}
an expression strictly holding in the limit $\nu\gg 1$, i.e., large masses.
For brevity we will refer to this prescription as `coagulation' rate. In
Fig.~10 we illustrate the creation, positive, formation, and coagulation
rates (color code) as a function of halo mass at different redshift, for an
ellipsoidal barrier; the former three rates are very similar for masses $M\ga
10^{10}\, M_\odot$ and redshift $z\ga 1$, with somewhat larger differences at
lower redshift and smaller masses. The coagulation rate is substantially
lower at small masses, although converges to the other prescriptions at large
masses (where actually Eq.~42 is supposed to hold).

\subsection{Halo growth history}

The halo conditional mass function can be exploited to construct a merger
tree, i.e., a numerical realization of the halo merger process. This
procedure with its delicate issues has been extensively discussed in the
literature (see Kauffmann \& White 1993; Somerville \& Kolatt 1999; Cole et
al. 2000; Parkinson et al. 2008; Neistein \& Dekel 2008; Zhang et al. 2008a),
and is beyond the scope of the present paper; however, from Eq.~(30) we can
derive the average growth history of DM halos, that would obtain after
averaging over many realization of a full merger tree. To this purpose, we
first compute the rate of accretion onto a mass $M$ at time $t$ as
\begin{equation}
\dot M(M,t) = \int_{M_{\rm min}}^M{\rm d}M'~(M-M')\, {{\rm d}^2
p_{M'\rightarrow M}\over {\rm d}M' {\rm d}t}~,
\end{equation}
and then integrate backward in time for $M(t)$ the first-order differential
equation provided by the relation $\dot M=\dot M(M[t],t)$.

To derive the `mean' evolutionary track of an individual halo we set the
lower integration limit $M_{\rm min}$ in Eq.~(43) to $0$ (see Miller et al.
2006, their Eq.~4); practically, a lower limit is provided by the `mass
resolution' of the tree. Note that, as shown by Neistein \& Dekel (2008),
there is no double-counting issue associated to the range of integration
since in the excursion set framework the typical merger event is not binary
but involves multiple progenitors or substantial mass accretion, even for
infinitesimal time-steps; this is mirrored in the integrand of Eq.~(43),
which is not symmetric with respect to $M'$ and $M-M'$.

Following Neistein et al. (2006, see their Appendix and Eqs. 31 and 33) we
can also compute a first-order, analytical estimate of the `main progenitor'
evolutionary track by setting $M_{\rm min}=M/2$; this corresponds to consider
at each branch of a merger tree only the most massive progenitor. The `mean'
and the `main progenitor' growth histories for different final masses at
$z=0$ are illustrated by the blue and red lines in the top panel of Fig.~11.
We stress that for the first time these outcomes have been obtained on
adopting the ellipsoidal barrier, and as such extend previous analysis based
on the constant barrier (e.g., Miller et al. 2006; Neistein et al. 2006) and
constitute novel results. The main progenitor history is plainly biased
toward early-times relative to the mean one, and can be considered the
typical history of halos residing in rich environments or onto peaks of the
initial density perturbation fields.

It is worth nociting that we recover the typical two-stage evolution of a
halo concurrently found in many state-of-the-art $N$-body simulations and
semi-analytic studies (see Zhao et al. 2003; Fakhouri et al. 2010; Wang et
al. 2011; Lapi \& Cavaliere 2011); such evolution comprises an early fast
collapse and a late slow accretion, roughly separated by the epoch when the
halo accretion timescale becomes comparable to the Hubble time. This
separation epoch is highlighted by the dots on each evolutionary track.

In Fig.~12 we compare our results with the growth histories extracted from
$N$-body simulations by McBride et al. (2009; see also earlier works by
Wechsler et al. 2002, van den Bosch 2002). These authors have shown that the
halo mass growth from simulations can be parameterized as $M(z)/M(0)=
(1+z)^\beta\, e^{-\gamma\, z}$ in terms of two shape parameters $\beta$ and
$\gamma$; for galactic halos with current mass $M(0)\approx 2-3\times
10^{12}\,M_\odot$ the average values $\beta=0.1$ and $\gamma=0.69$ apply. The
corresponding evolution is illustrated in the Figure by crosses. Our
reference result for the growth history of a halo with mass $2.5\times
10^{12}\,M_\odot$ is reported as a cyan line. However, for fair comparison
with McBride et al. (2009), we must redo the computation after applying two
changes: first, we include in the minimum mass $M_{\rm min}$ entering
Eq.~(43) their mass resolution of $10^9\, M_{\odot}$ (green line); second, we
modify our normalization of the mass variance $\sigma_8=0.81$ to match their
value $0.9$ (orange line). Our overall outcome is in striking agreement with
the $N$-body result.

The bottom panel of Fig.~11 is focused on evolutionary tracks starting at
$z=2$ instead of $z=0$. Comparing the two panels, it is evident that the
histories ending at $z=2$ feature a more rapid evolution; for example, halos
of $10^{13}\, M_\odot$ at $z=2$ are still in, or very close to, their fast
collapse stage, while the same halos at $z=0$ are well within the slow
accretion regime. This implies that the modeling of the baryonic processes
occurring inside a given halo should be quite different, depending on whether
its formation occurred at low or high redshift.

In the vein of a link to observations, we remark that such massive halos at
$z\sim 2$ are the host of bright submillimeter (submm) selected galaxies, as
demonstrated by studies of the clustering properties and of the cosmic
infrared background autocorrelation function (see Cooray et al. 2010; Xia et
al. 2012). Specifically, Lapi et al. (2011) have shown that these galaxies
selected with limiting flux $S\ga 35$ mJy at $250\,\mu$m exhibit star
formation rates in the range $300-3000\, M_{\odot}$ yr$^{-1}$. This implies
that the fast collapse of massive halos at these substantial redshift $z\ga
1.5$ originates physical conditions in the associated baryons leading to
rapid cooling and condensation into stars. As expected from basic physical
arguments and shown by numerical simulations (e.g., Zhao et al. 2003, Wang et
al. 2011), the halo fast collapse includes the rapid merging of large clumps,
losing most of their angular momentum by relaxation processes over a few
dynamical times (see Lapi \& Cavaliere 2011). This concurs with high
resolution photometric and spectroscopic observations of $z\sim 2$ galaxies,
showing that the star formation occurs in clumps located within the central
kpc-sized regions, and that any residual rotational motion is unstable and
expected to be dissipated over a few hundred million years (e.g., Genzel et
al. 2011). On the other hand, during the slow accretion stage at low
redshift, a quite different baryonic evolution is expected.

\subsection{Halo bias}

The excursion set approach may be exploited to evaluate the correlation
between halo abundances and their environment. This is usually done by
computing the (Eulerian) halo bias (see Mo \& White 1996; Sheth \& Tormen
1999)
\begin{equation}
b(M,z)= 1+{1\over \delta_{c0}}\,\left[{N(M, \delta_c\rightarrow
M_0,\delta_{c0})\over N(M,\delta_c)\, V}-1\right]~;
\end{equation}
in the expression above what matters is the ratio between the number of halos
at redshift $z$ with density contrast $\delta_c$ that will end up in an
environment with volume $V$, mass $M_0\simeq \bar\rho V\gg M$, and density
contrast $\delta_{c0}\ll \delta_c$; we have denoted ${\rm d}N/{\rm d}M$ with
$N$ to simplify the notation.

Since conditions where $\Delta S\simeq S\gg S_0$ are relevant in this
context, we cannot exploit our approximate expression for the conditional
mass function derived in \S~4; however, it is rather easy to single out that,
when written in terms of the appropriate scaling variable $\nu_c\equiv
(\Delta \delta_c)^2/\Delta S$ its shape should be close to the unconditional
distribution; this is because the barrier of the conditional problem
$B(\Delta S, \delta_c, \delta_{c0})=B(S,\delta_c)-B(S_0, \delta_{c0})\simeq
B(S, \Delta \delta_c)$ tend to the unconditional one as $S\gg S_0$. Using
Eqs.~(21) and (44) one obtains
\begin{equation}
b(M,z)= 1+{1\over \delta_{c0}}\,\left[{\nu_c\, f(\nu_c)\over \nu\,
f(\nu)}-1\right]\simeq 1-{2\over \delta_c}\,\left[1+\,{{\rm d}\log f\over
{\rm d}\log\nu}\right]~.
\end{equation}
where $f$ is the unconditional distribution; the second equality follows from
considering that $\nu_c\simeq \nu (1-2\delta_{c0}/\delta_c)$ and
$f(\nu_c)\simeq f(\nu)-2\, (\delta_{c0}/\delta_c)\,\nu f'(\nu)$ in the
relevant limits $S\gg S_0$ and $\delta_c\gg \delta_{c0}$. Note that in this
way we obtain the bias function for halos observed at redshift $z_{\rm
obs}=0$ but formed at redshift $z>z_{\rm obs}$; on the other hand, the bias
for halos observed as soon they are formed at $z_{\rm obs}=z$ can be obtained
by replacing $\delta_c(z)\rightarrow \delta_c(z_{\rm obs})$ in Eq.~(45).

Our results of using the explicit expression of $f(\nu)$ from Eq.~(21) into
Eq.~(45) are illustrated in Fig.~13 as a function of halo mass at different
redshifts, for the constant, square-root and ellipsoidal barriers
(color-coded); top panel is for $z_{\rm obs}=z$ and bottom panel is for
$z_{\rm obs}=0$. Our results for the ellipsoidal barrier compare well to the
outcomes of cosmological $N$-body simulations (crosses; fit to Tinker et al.
2005), although tend to overpredict the asymptotic bias of low-mass halos and
slightly underpredict that of high-mass halos (see also discussion by Tinker
et al. 2010). We stress that, as explicitly shown by Ma et al. (2011, see
their Table~1 and Fig.~1) for the case of a constant barrier, the inclusion
of non-Markovian effects and of stochasticity in the barrier can be relevant
in obtaining a better agreement between the bias function computed from the
excursion set approach with that measured from $N$-body simulations.

\section{Summary and conclusions}

The formation and evolution of dark matter halos constitutes a highly complex
problem, whose attack ultimately requires cosmological $N$-body simulations
on supercomputers. However, some analytic grasp is most welcome to better
interpret their outcomes, to provide approximated yet flexible analytic
representations of the results, to develop strategies for future simulation
setups, and to quickly explore the effects of modifying the excursion set
assumptions or the cosmological framework (\S~1).

To this purpose, we have exploited the excursion set approach in integral
formulation (\S~2) to derive novel, accurate approximated expressions of the
unconditional (\S~3) and conditional (\S~4) first crossing distributions, for
random walks with uncorrelated steps and general shapes of the barrier. We
find the corresponding approximation of the halo mass functions for Cold Dark
Matter power spectra to represent very well the outcomes of state-of-the-art
cosmological $N$-body simulations. In addition, we have apllied these results
to derive and confront with simulations (\S~5) various quantities of interest
in halo statistics, including the rates of halo formation and creation, the
average halo growth history, and the halo bias for general moving barriers.

In Table~1 we list our main results, recall their location in the paper, and
cross-reference to the corresponding equations and figures. A library of
IDL-based routines to easily compute the quantities derived in this paper is
available at the URL \texttt{http://people.sissa.it/$\sim$ lapi/rates}
(currently under construction).

Finally, we have discussed how our results are affected when considering
random walks with correlated instead of uncorrelated steps (Appendix B), and
Warm instead of Cold Dark Matter power spectra (Appendix C). We stress that
above and around the Warm Dark Matter free-streaming mass scale, our
approximations to the halo mass function agree very well with the outcomes of
current $N-$body simulations.

\begin{acknowledgements}
Work supported in part by ASI, INAF and MIUR. We thank the referee for
constructive comments and helpful suggestions. We acknowledge stimulating
discussions with A. Cavaliere, P.S. Corasaniti, G. De Zotti, S. Liberati, J.
Moreno, M. Musso, F. Shankar, and G. Pradisi. AL thanks SISSA for warm
hospitality.
\end{acknowledgements}

\begin{appendix}

\section{$\zeta$-regularization technique for divergent integrals}

Here we provide a brief primer on the $\zeta$-regularization technique for
divergent series and integrals, helpful in \S~5.1; the interested reader can
find more mathematical details and physical applications in classic textbooks
like Hardy (1949), Birrell \& Davies (1984), and Elizalde et al. (1994).

The $\zeta$-function was originally defined by Euler as
\begin{equation}
\zeta(s)=\sum_{n=1}^\infty{1\over n^s}~;
\end{equation}
this series is convergent for any complex number $s$ with
$\mathfrak{Re}(s)>1$, i.e., real part exceeding $1$.

Riemann suggested to extend the definition of $\zeta(s)$ to the whole complex
plane by analytic continuation. For example, one can use the expression
involving the Dirichlet alternating series
\begin{equation}
\zeta(s)={1\over 1-2^{-s}}\, \sum_{n=1}^\infty{(-1)^{n-1}\over n^s}~
\end{equation}
which coincides with the standard definition above for $\mathfrak{Re}(s)>1$
but is well defined for any $s$ with $\mathfrak{Re}(s)>0$ except the point
$s=1$ which is a single pole. For $s\simeq 1$ it can be shown that $\zeta(s)$
features the approximate behavior
\begin{equation}
\zeta(s)\simeq {1\over s-1}+\gamma_{\rm em}~,
\end{equation}
where $\gamma_{\rm em}=0.57721...$ is the Euler-Mascheroni constant.

In addition, Riemann discovered that the $\zeta$-function satisfies the
functional equation
\begin{equation}
\zeta(s)=2^s \pi^{s-1}\,\sin\left({\pi s\over 2}\right)\,\Gamma(1-s)\,
\zeta(1-s)~,
\end{equation}
where $\Gamma(s)\equiv \int_0^\infty{\rm d}t\, t^{s-1}\, e^{-t}$ is the Euler
$\Gamma$-function. This can be used to extend the definition of $\zeta(s)$
even to values of $s$ with $\mathfrak{Re}(s)\leq 0$. In particular, for
$s\rightarrow 0$ the functional equation yields
\begin{equation}
\zeta(0)\simeq {1\over \pi}\,{\pi s\over 2}\, {1\over (1-s)-1} = -{1\over 2}~.
\end{equation}
This is the origin of the weird-looking expression
\begin{equation}
\zeta(0) = \sum_{n=1}^\infty {1\over n^0} = 1+1+1+... = -{1\over 2}~;
\end{equation}
in fact, the reader should keep in mind that $\zeta(s)$ for $s=0$ is only
\emph{formally} related to the infinite sum at the second and third member.

The value $\zeta(0)$ derived above can be effectively used to regularize the
divergent improper integral
\begin{equation}
\lim_{\Lambda\rightarrow \infty} \int_0^{\Lambda}{\rm
d}x=\lim_{\Lambda\rightarrow \infty} \Lambda=\infty~;
\end{equation}
this can be done by replacing the integral with a series, that turns out to
be closely related to $\zeta(0)$, in the form
\begin{equation}
\lim_{\Lambda\rightarrow \infty} \int_0^{\Lambda}{\rm
d}x=\lim_{\Lambda\rightarrow \infty} \sum_{n=0}^{\Lambda}\, 1=
1+\lim_{\Lambda\rightarrow \infty} \sum_{n=1}^{\Lambda}\, {1\over
n^0}=1+\zeta(0)=1-{1\over 2}={1\over 2}~.
\end{equation}

Integrals involving more general integrands but with the same diverging
behavior can also be easily regularized by adding and subtracting convenient
quantities and using the result of Eq.~(A8). For example, we show below how
to regularize the divergent integral appearing in \S~5.2 ($\eta$ is a
positive constant):
\begin{eqnarray}
\nonumber \lim_{\Lambda\rightarrow \infty} \int_0^{\eta\Lambda}{\rm
d}x~e^{-k^2/2
x^2}&=&\lim_{\Lambda\rightarrow\infty}\left\{\int_0^{\eta\Lambda}{\rm
d}x~e^{-k^2/2 x^2}-\eta\Lambda+\eta\int_0^{\Lambda}{\rm d}x\right\}=\\
&& \nonumber \\
&=&\lim_{\Lambda\rightarrow\infty}\left\{\left[x\,e^{-k^2/2
x^2}+\sqrt{\pi\over 2}\,k\,\mathrm{erf}\left({k\over \sqrt{2}
x}\right)\right]_0^{\eta\Lambda}-\eta\Lambda\right\}+{\eta\over
2}=\\
&& \nonumber\\
\nonumber &=& \eta\Lambda-\sqrt{\pi\over 2}\, k-\eta\Lambda+{\eta\over
2}=-\sqrt{\pi\over 2}\, k+{\eta\over 2}~.
\end{eqnarray}

\section{Excursion set approach with correlated steps}

In the main text we have been concerned with the first crossing problem when
the steps of the random walk are uncorrelated. In this Appendix we discuss
how the integral formulation of the excursion set approach introduced in \S~2
can be modified to account for some degree of correlation between the steps.
Recently, correlated random walks have received considerable interest (see
Paranjape et al. 2012; Musso \& Sheth 2012); in fact, it may well be that
such models are more realistic. This is because the random walk $\delta(S)$
is truly Brownian, i.e., with uncorrelated steps, only if a sharp filter
function in Fourier-space is adopted. However, as pointed out in \S~2 in this
case there is an ambiguity of a sort in associating a mass to the filtering
scale. On the other hand, using a real-space top-hat window function gives a
well-defined relation between mass and smoothing scale, but necessarily
introduces some degree of correlations among the steps of the random walk.
Note that for the slowly-varying Cold DM power spectra, this problem is
somewhat academic, since the shape of $S(M)$ depends very weakly on the
choice of the filter function; however, it is not so for, e.g., Warm DM
spectra (see discussion by Benson et al. 2013). In addition, correlations
between the steps of the walk can also be introduced by non-Gaussian features
in the initial spectrum of perturbations (see Maggiore \& Riotto 2010c).

For the sake of definiteness, here we mainly consider the extreme instance
according to which the steps of the random walk are \emph{completely}
correlated. Then one expects that the walk $\delta(S)$ does not proceed
through stochastic zigzags, but instead grows almost monotonically with $S$.
This means that if $\delta$ first exceeded the barrier $B(S)$ at $S$, it was
certainly below that at all $S'<S$. Then the constraints imposed by Eq.~(12)
is superfluous and $P(\delta,S)\simeq P_0(\delta,S)$ holds for $\delta<B(S)$
to a very good approximation. Simple algebra yields
\begin{equation}
\int_0^S{\rm d}S'~f(S') = {1\over 2}\, \mathrm{erfc}\left[{B(S)\over \sqrt{2
S}}\right]~,
\end{equation}
which constitutes the counterpart of Eq.~(13) for completely correlated
steps.

Differentiating with respect to $S$ leads to the first crossing distribution
\begin{equation}
f(S)=-{e^{-B^2(S)/2 S}\over \sqrt{2\pi}}\, {{\rm d}\over {\rm d}S}
\left[{B(S)\over \sqrt{S}}\right]~.
\end{equation}
Specializing to the barrier shape $B(S)=B_0+B_\gamma\, S^\gamma$ yields
\begin{equation}
f(S)={B_0\over 2}\, {e^{-B^2(S)/2 S}\over \sqrt{2\pi S^3}}\,
\left[1+(1-2\,\gamma)\, {B_\gamma\over B_0}\, S^\gamma\right]~.
\end{equation}
Note that in the constant barrier case ($B_\gamma=0$), this is exactly half
of the result for uncorrelated steps, and actually coincides with the
expression found by Press \& Schechter (1974).

In Fig.~14 we compare the unconditional first crossing distribution for
uncorrelated (solid lines) and completely-correlated (dashed lines) steps. It
is apparent that the outcome of cosmological $N$-body simulations (crosses)
lies much closer to the excursion set result for uncorrelated steps. This is
also the case if one consider strongly but not completely correlated steps
like in the framework developed by Musso \& Sheth (2012), whose results are
illustrated in Fig.~14 by the dotted lines.

All in all, models with correlated steps need much of an improvement before
reproducing the halo mass function at a level comparable to models with
uncorrelated steps; this improvement, although only on an effective/empirical
basis, may come from considering a stochastic in place of a deterministic
barrier possibly including some drift (see Robertson et al. 2009; Maggiore \&
Riotto 2010b; Corasaniti \& Achitouv 2011).

\section{Excursion set approach with Warm Dark Matter}

In the main text we have performed computations for Cold Dark Matter as a
reference. In this Appendix we show that our formulation of the excursion set
approach is extremely flexible with respect to changes of the power spectrum.
Specifically, here we focus on Warm Dark Matter (e.g., Dodelson \& Widrow
1994), and test our approximation to the halo mass function against recent
$N$-body simulations.

To this purpose, we follow Bode et al. (2001) and Barkana et al. (2001) in
modifying the Cold DM power spectrum to impose a cutoff below the Warm DM
free-streaming length scale, and in including the mass-dependent behavior of
the linear threshold for collapse $\delta_c(z,M)$ enforced by the Warm DM
particle's residual thermal velocities. Specifically, for the resulting power
spectrum we take Eqs.~(3) and (4) in Barkana et al. (2001), while for the
threshold we take Eqs.~(7) to (10) in Benson et al. (2013).

Practically, in computing the mass function after Eq.~(9) we exploit our
results on the first crossing distribution for an ellipsoidal barrier
Eq.~(20), then connect variance $S$ and mass $M$ with the Warm DM power
spectrum, and finally introduce the mass-dependent $\delta_c(M,z)$ in the
result a posteriori. This procedure provides reasonable results above the
free-streaming mass scale. In fact, in Fig.~15 we compare our result on the
mass function at $z=0$ for $m_{\rm FD}=0.25$ keV with the outcomes from the
recent $N-$body simulation by Schneider et al. (2012), finding an excellent
agreement. However, below the free-streaming mass scale the computations
based on the ellipsoidal barrier are quite uncertain, because the barrier
$B(S,t)$ itself is poorly understood. In fact, there is some debate as to
whether halos with such small masses can form at all (see Bode et al. 2001;
Wang \& White 2007). If this is the case, the overall result would be a more
pronounced cutoff of the mass function just below the free-streaming scale
(see discussion by Smith \& Markovic 2011; Menci et al. 2012; Benson et al.
2013; also the recent simulations by Schneider et al. 2013). To highlight
this uncertainty, in Fig.~15 we have indicated with a thick line our results
above, and with a thin line below, the free-streaming scale.

In Fig.~15 we also present the outcomes of the unconditional mass function at
different redshifts for values of the Warm DM particle mass (color-coded)
$m_{\rm FD}= 0.25$, $0.75$ and $1.5$ keVs, compared with the standard Cold DM
outcome. As customary we quote here the Fermi-Dirac mass $m_{\rm FD}$, i.e.,
the mass that the particles would have if they were thermal relics (decoupled
in thermal equilibrium); this is convenient because the masses of Warm DM
particles produced in different microscopic scenarios can be easily related
to $m_{\rm FD}$, see de Vega \& Sanchez (2012).

All in all, the Warm DM results are practically indistinguishable from the
Cold DM ones at the high mass end, then on approach the free-streaming mass
progressively deviate downward (and then are practically cut-off). The masses
where the deviation sets in are smaller for lower Fermi-Dirac mass, and
higher redshift. In particular, we stress that for $m_{\rm FD}\la 1.5$ keV,
at redshift $z\ga 6$ and masses $M\la 10^{11}\, M_\odot$ the strongly
flattening behavior of the halo mass function should be mirrored by the faint
end in the luminosity function of Lyman Break Galaxies up to $z\sim 10$
(e.g., Bouwens et al. 2011); there dust obscuration should not constitute a
relevant issue, and the flattening could be so prominent not to be easily
ascribed to supernova feedback (see Mao et al. 2007; Cai et al. 2013, in
preparation).

\end{appendix}

\clearpage
\begin{figure}
\epsscale{1.0}\plotone{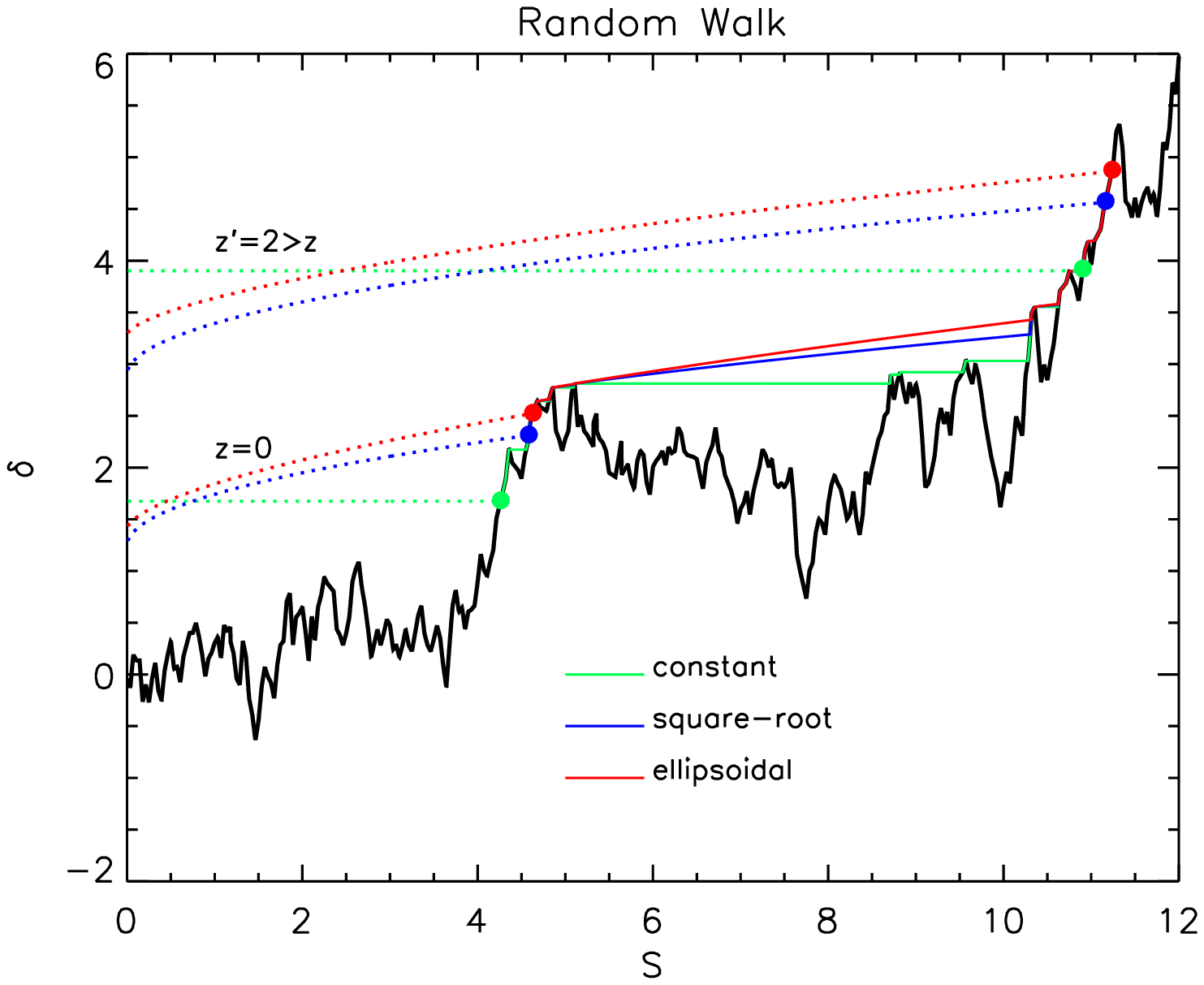}\caption{Example of a random walk executed by
the overdensity contrast $\delta$ as a function of the variance $S$. The
black solid line represents the random walk. The dotted lines illustrate the
constant (green), square-root (blue), and ellipsoidal barriers (red) at two
redshifts $z=0$ and $z'=2>z$, with the dots indicating the locations of first
crossing. Moving toward the present, the barrier heights decreases and the
points of first crossing occur at smaller $S$, corresponding to the formation
of a more massive halo by merger. This growth history is represented by the
solid colored lines connecting steps of the random walk curve; the longer the
line between two steps, the more massive the merger with the companion halo.}
\end{figure}

\clearpage
\begin{figure}
\epsscale{1.0}\plotone{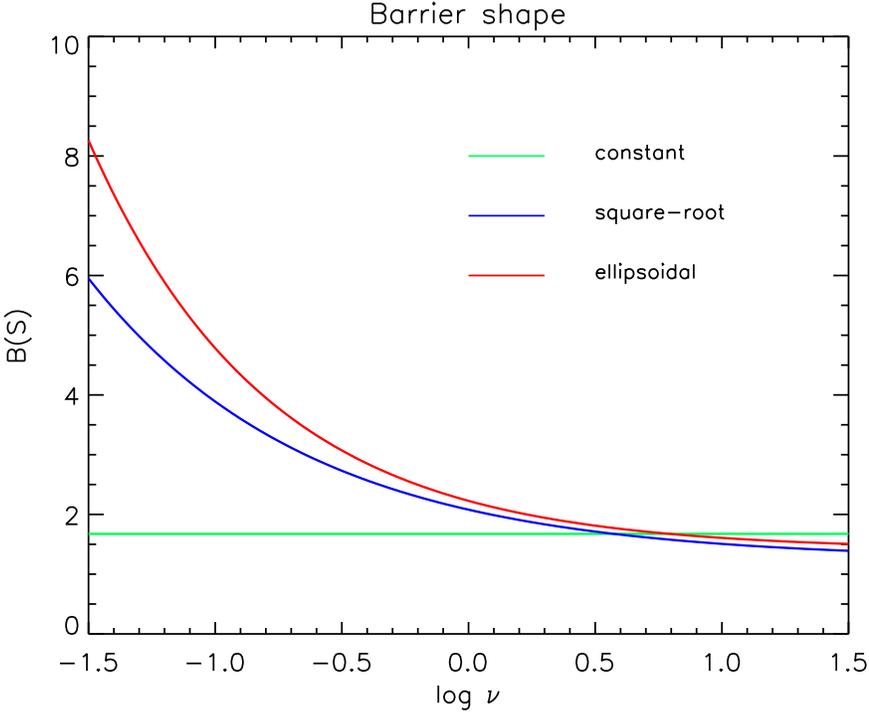}\caption{Barrier shapes considered as
prototypical examples in this work: constant (green), square-root (blue), and
ellipsoidal (red). For their analytic expressions see \S~2 of the main text.}
\end{figure}

\clearpage
\begin{figure}
\epsscale{1.0}\plotone{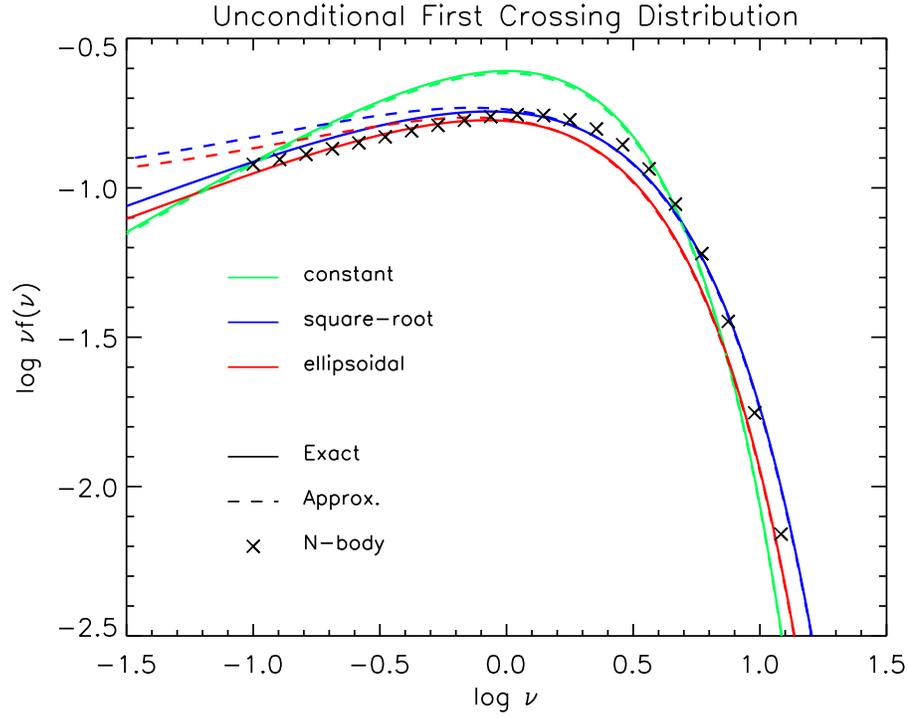}\caption{Unconditional first crossing
distribution, as a function of the self-similar variable $\nu\equiv
\delta_c^2/S$, for the constant (red), square-root (blue), and ellipsoidal
(red lines) barriers. Solid lines refer to the exact results on solving the
integral Eq.~(13) via Eq.~(16), dashed lines illustrate our approximation
given by Eq.~(21), and crosses show the outcomes of cosmological $N$-body
simulations (fit to Tinker et al. 2008).}
\end{figure}

\clearpage
\begin{figure}
\epsscale{1.0}\plotone{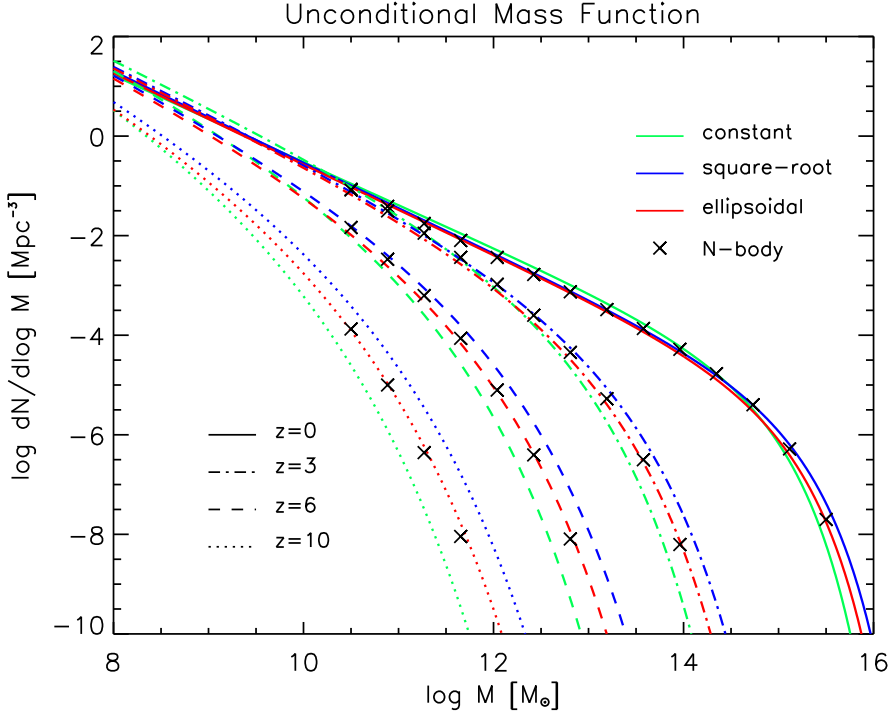}\caption{Unconditional mass function as a
function of halo mass at different redshifts, for the constant (green),
square-root (blue), and ellipsoidal (red) barriers. Crosses illustrate the
the outcomes of $N$-body simulations (fit to Tinker et al. 2008). A standard
Cold DM power spectrum has been adopted.}
\end{figure}

\clearpage
\begin{figure}
\epsscale{1.0}\plotone{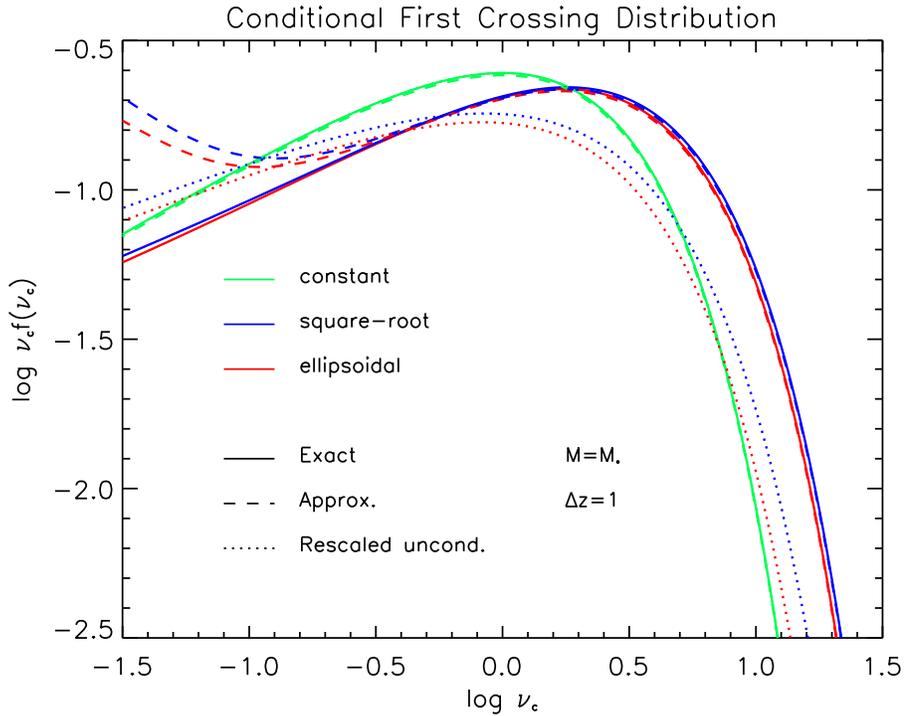} \caption{Conditional first crossing
distribution, as a function of the self-similar variable $\nu_c\equiv (\Delta
\delta_c)^2/\Delta S$, for the constant (red), square-root (blue), and
ellipsoidal (red lines) barriers. Solid lines refer to the exact results on
solving the integral Eq.~(13) via Eq.~(16), dashed lines illustrate our
approximation given by Eq.~(28), and dotted lines refer to the rescaled
unconditional distribution rescaled to $\nu_c$. The figure refers to the a
descendent mass $M\approx 10^{13}\, M_\odot$ (close to the characteristic
mass at $z=0$) and to a difference $\Delta z=1$ between the redshift of the
progenitors and of the descendant. A standard Cold DM power spectra has been
adopted.}
\end{figure}

\clearpage
\begin{figure}
\epsscale{1.0}\plotone{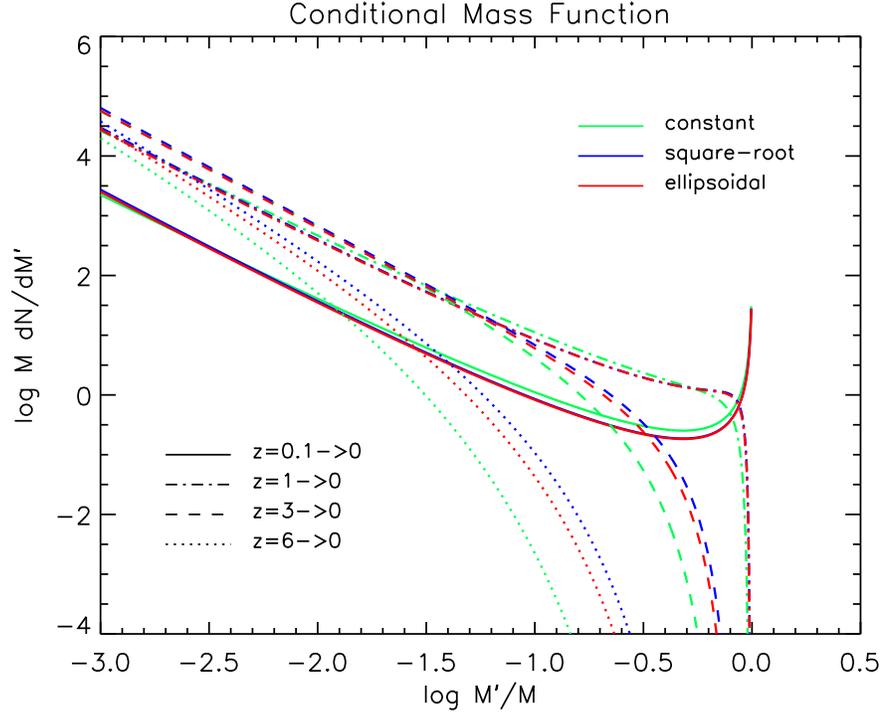} \caption{Conditional mass function as a
function of progenitor halo mass, for a descendent mass of $10^{13}\,
M_\odot$ and various differences between redshift of the progenitors and the
descendant. Color code refers to the constant (green), square-root (blue),
and ellipsoidal (red) barriers. A standard Cold DM power spectra has been
adopted.}
\end{figure}

\clearpage
\begin{figure}
\epsscale{1.0}\plotone{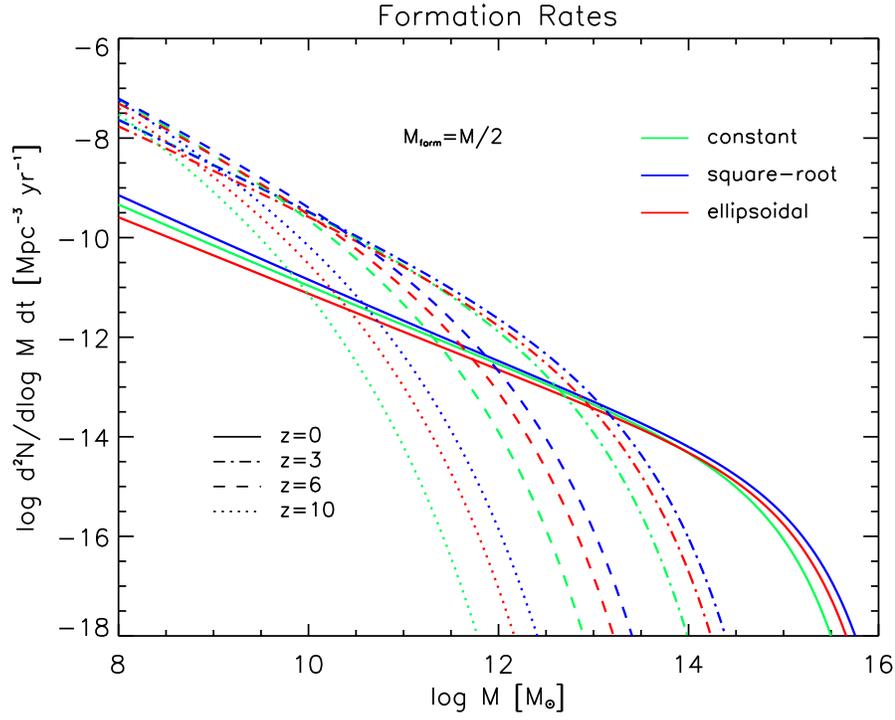}\caption{Formation rates as a function of halo
mass at different redshifts, for the constant (green), square-root (blue) and
ellipsoidal (red) barriers. A formation mass $M_{\rm form}=M/2$ and a
standard Cold DM power spectra have been adopted.}
\end{figure}

\clearpage
\begin{figure}
\epsscale{1.0}\plotone{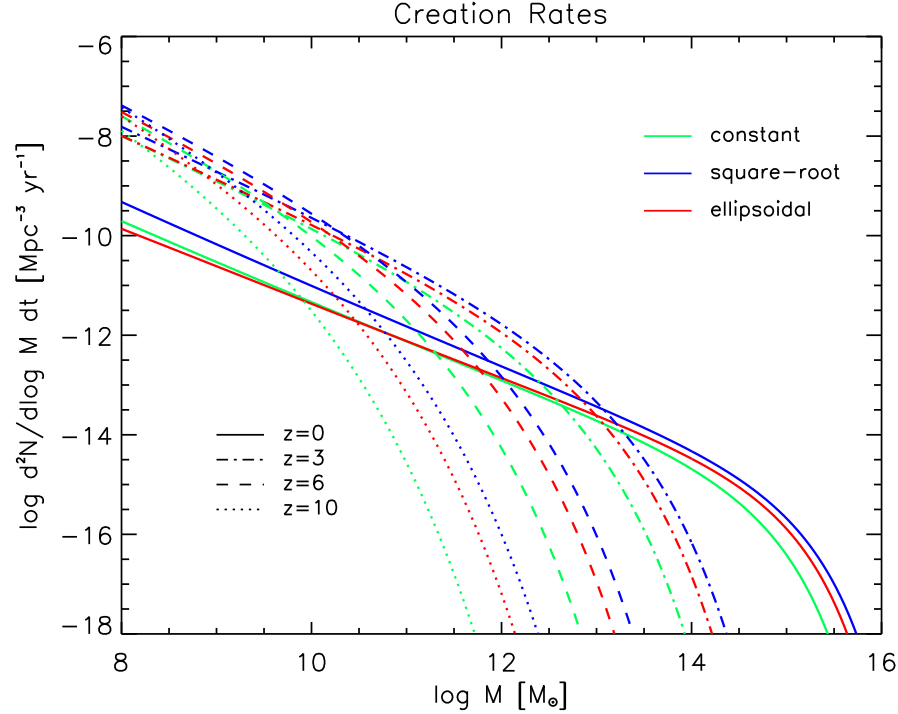}\caption{Creation rates as a function of halo
mass at different redshifts, for the constant (green), square-root (blue) and
ellipsoidal (red) barriers. A standard Cold DM power spectra has been adopted.}
\end{figure}

\clearpage
\begin{figure}
\epsscale{1.0}\plotone{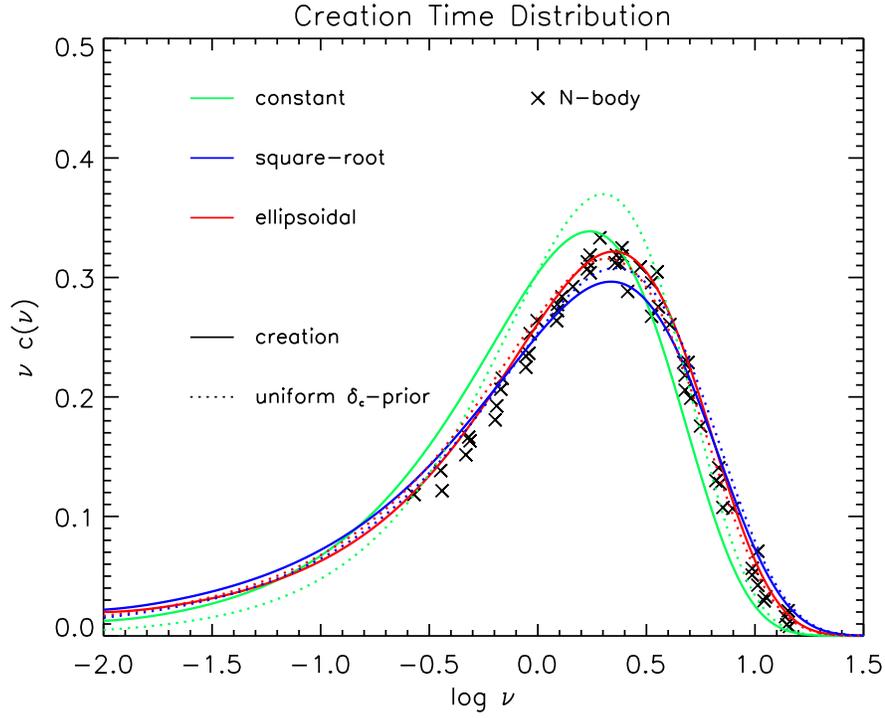}\caption{Creation time distribution as a
function of the self-similar variable $\nu\equiv \delta_c^2/S$ for the
constant (green), square-root (blue) and ellipsoidal (red) barriers. Solid
lines refer to our prescription for the creation rates, while dotted lines
refer to the procedure envisaged by Percival \& Miller (1999), crosses are
the results of $N$-body cosmological simulations (Moreno et al. 2009). A
standard Cold DM power spectra has been adopted.}
\end{figure}

\clearpage
\begin{figure}
\epsscale{1.0}\plotone{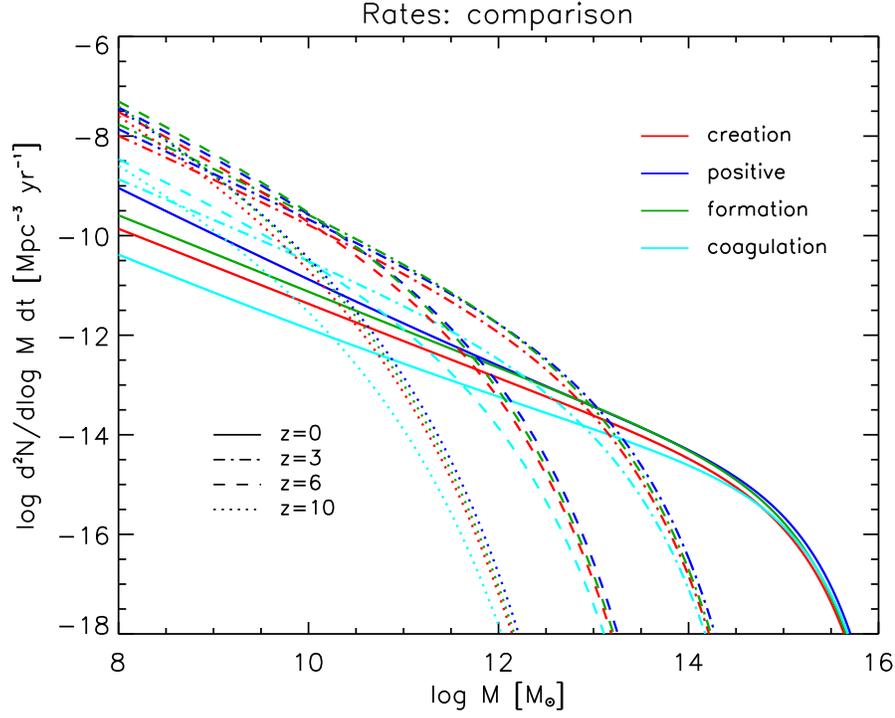}\caption{Comparison between different
prescriptions to compute halo rates, as a function of halo mass at
different redshifts: creation rates (red) computed according Eq.~(38),
positive time derivative of unconditional mass function (blue) according
Eq.~(41), formation rates (green) according Eq.~(34), coagulation rates
(cyan) according Eq.~(42). An ellipsoidal barrier and a standard Cold DM power
spectra have been adopted.}
\end{figure}

\clearpage
\begin{figure}
\epsscale{0.8}\plotone{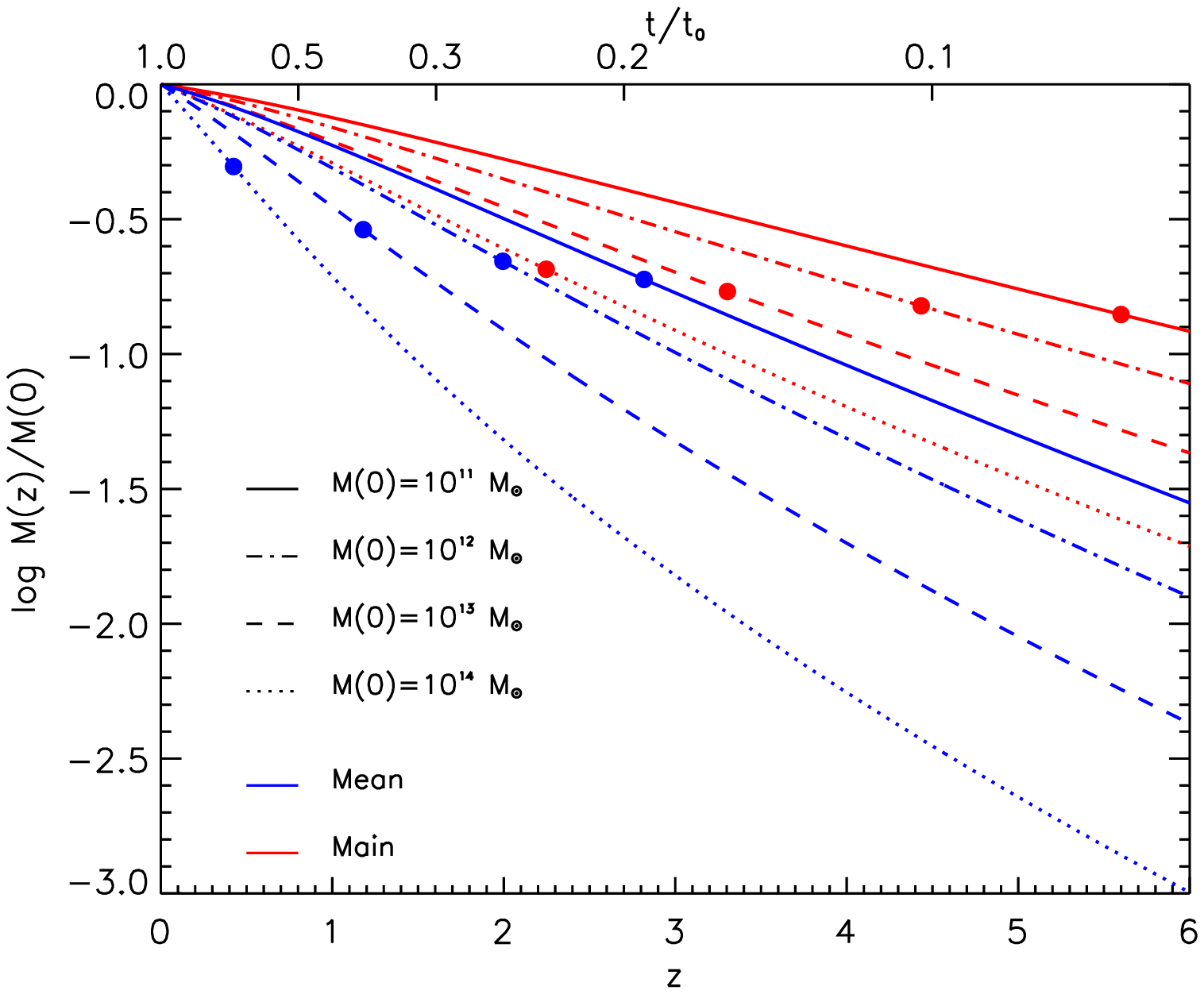}\plotone{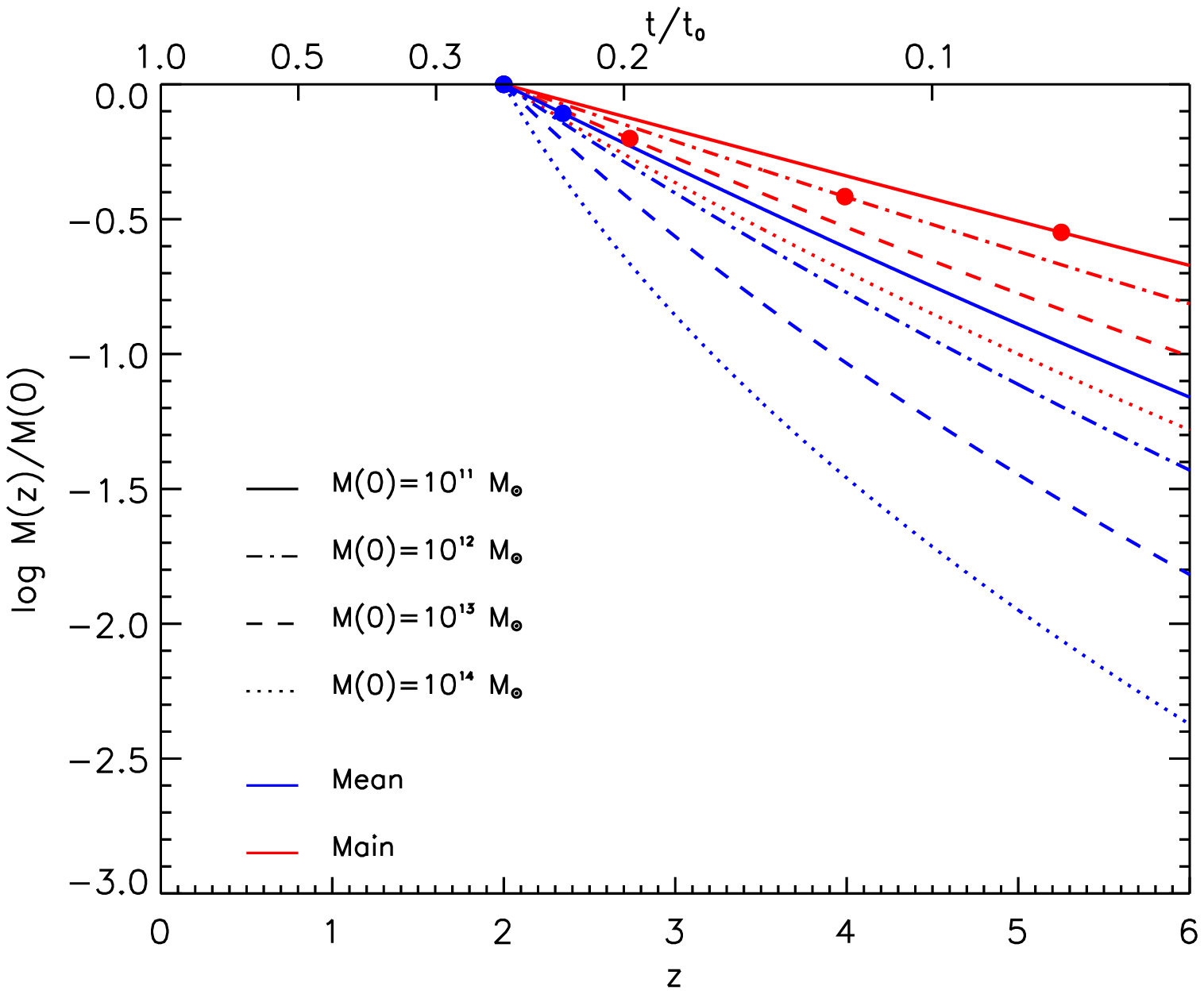}
\caption{Mass growth history of DM halos as a function of redshift (lower
scale; the upper scale refers to the cosmological time in units of the
present age of the Universe $t_0=13.7$ Gyr), for different masses $10^{11}$
(solid), $10^{12}$ (dot-dashed), $10^{13}$ (dashed), $10^{14}\, M_\odot$
(dotted) at $z=0$ (top panel) and at $z=2$ (bottom panel). Blue lines refer
to the mean history, and red lines to the main progenitor one (see \S~5.2
for details). The dots illustrate the epoch separating the early fast
collapse from the late slow accretion, i.e., when the typical accretion
timescale becomes comparable to the Hubble time (in the bottom panel only the
smaller masses have met this condition). The ellipsoidal barrier and a
standard Cold DM power spectrum have been adopted.}
\end{figure}

\clearpage
\begin{figure}
\epsscale{1.0}\plotone{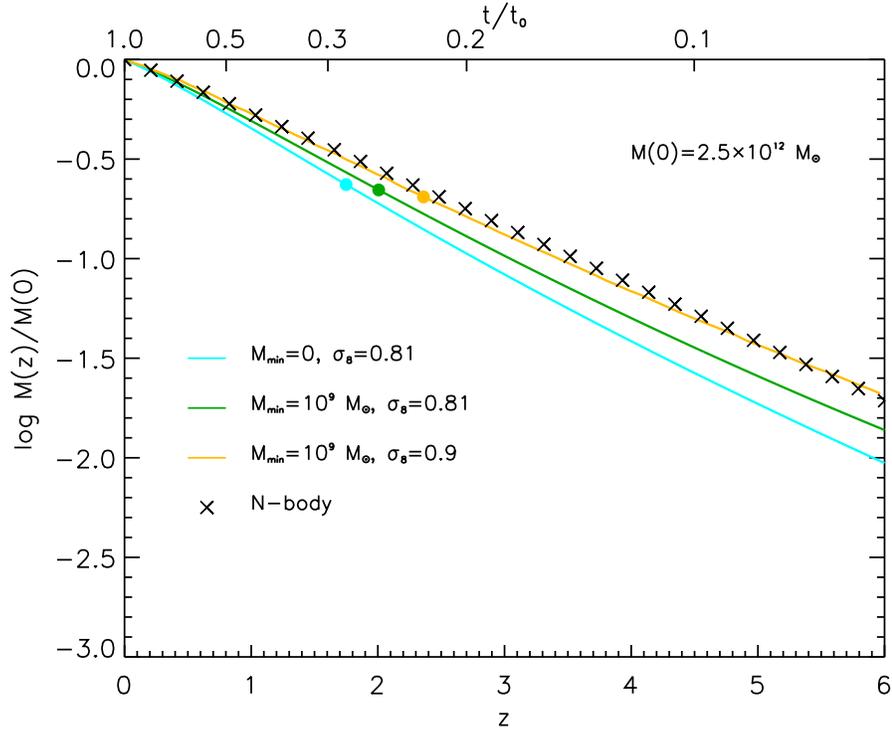} \caption{Comparison between our
computation based on Eq.~(43; colored lines) and the outcomes of $N$-body
simulations (see McBride et al. 2009; crosses) for the growth history of a DM
halo with current mass $M(0)=2.5\times 10^{12}\, M_\odot$. Cyan line is our
result for $M_{\rm min}=0$ in Eq.~(43) corresponding to an infinite mass
resolution, and for the value $\sigma_8=0.81$ of the mass variance
normalization; green line is for $M_{\rm min}=10^{9}\, M_{\odot}$ but still
$\sigma_8=0.81$; orange line is for $M_{\rm min}=10^{9}\, M_{\odot}$ and
$\sigma_8=0.9$, to match the values adopted in the $N-$body simulations
analyzed by McBride et al. (2009).}
\end{figure}

\clearpage
\begin{figure}
\epsscale{0.8}\plotone{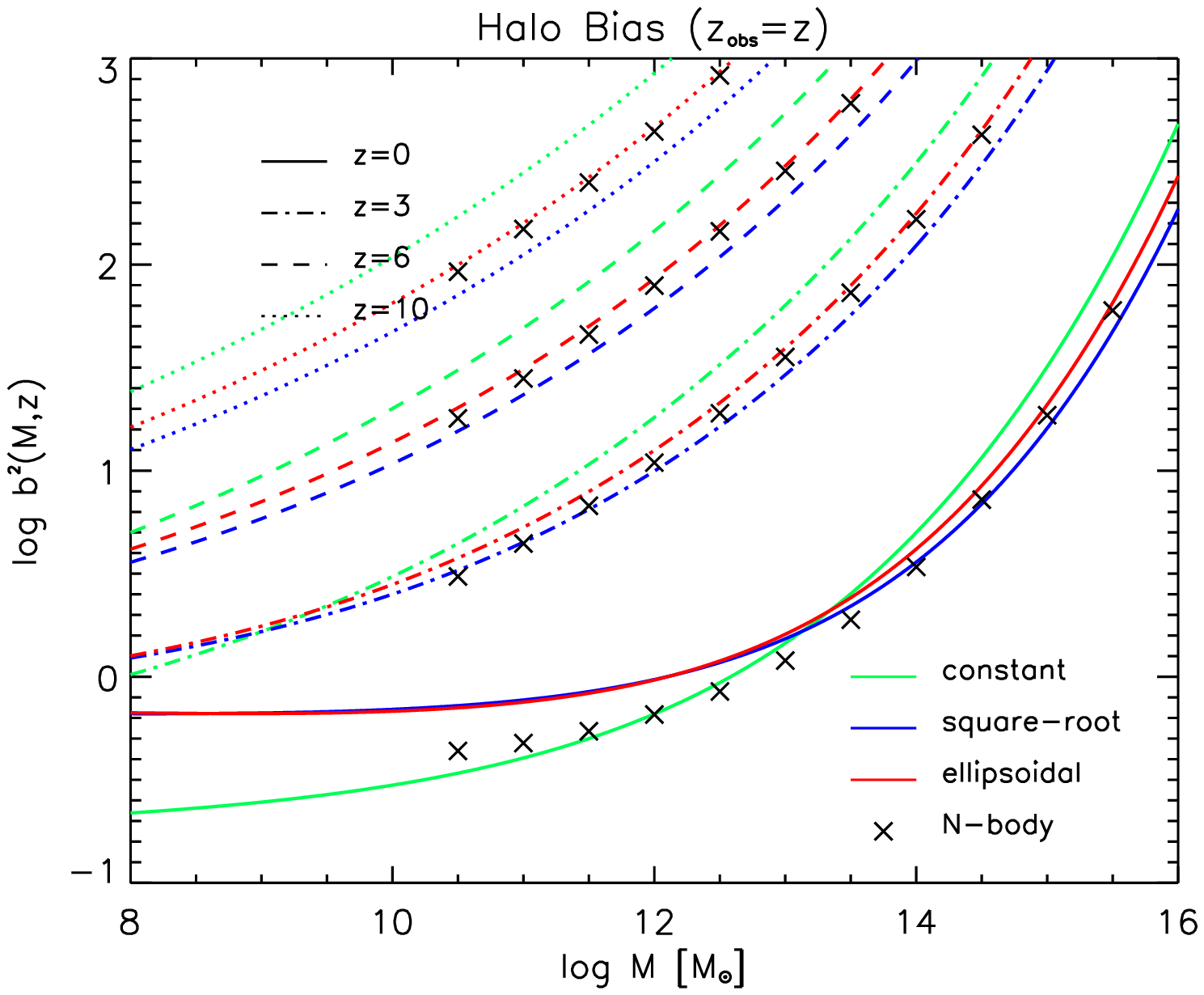}\\\plotone{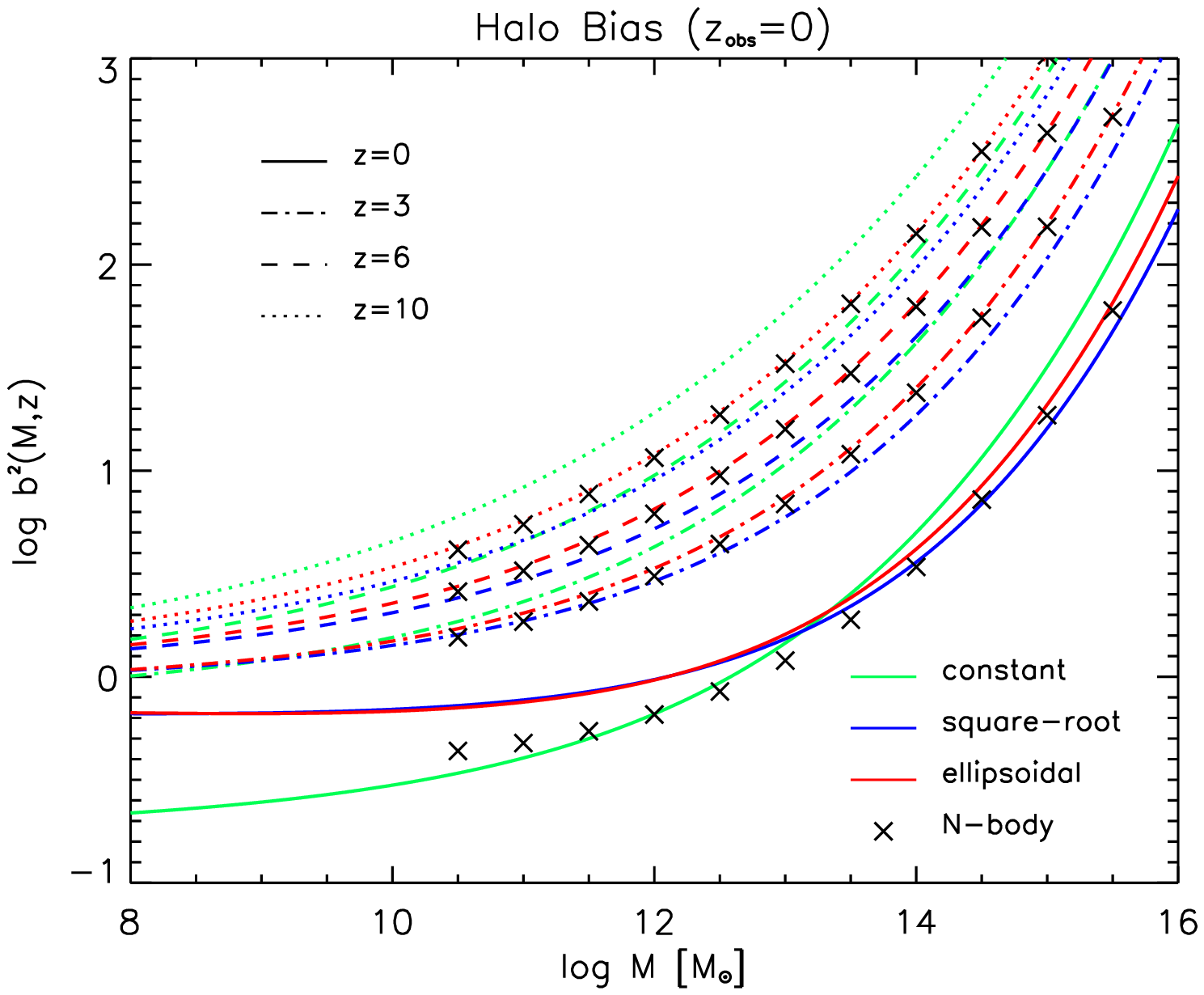}\caption{Halo bias as a
function of halo mass at different formation redshift $z$, for the constant
(green), square-root (blue) and ellipsoidal (red) barriers. Top panel is for
observation redshift $z_{\rm obs}=z$ and bottom one for $z_{\rm obs}=0$.
Crosses are the results of $N$-body cosmological simulations (fit to
Tinker et al. 2005). A standard Cold DM power spectra has been adopted.}
\end{figure}

\clearpage
\begin{figure}
\epsscale{1.0}\plotone{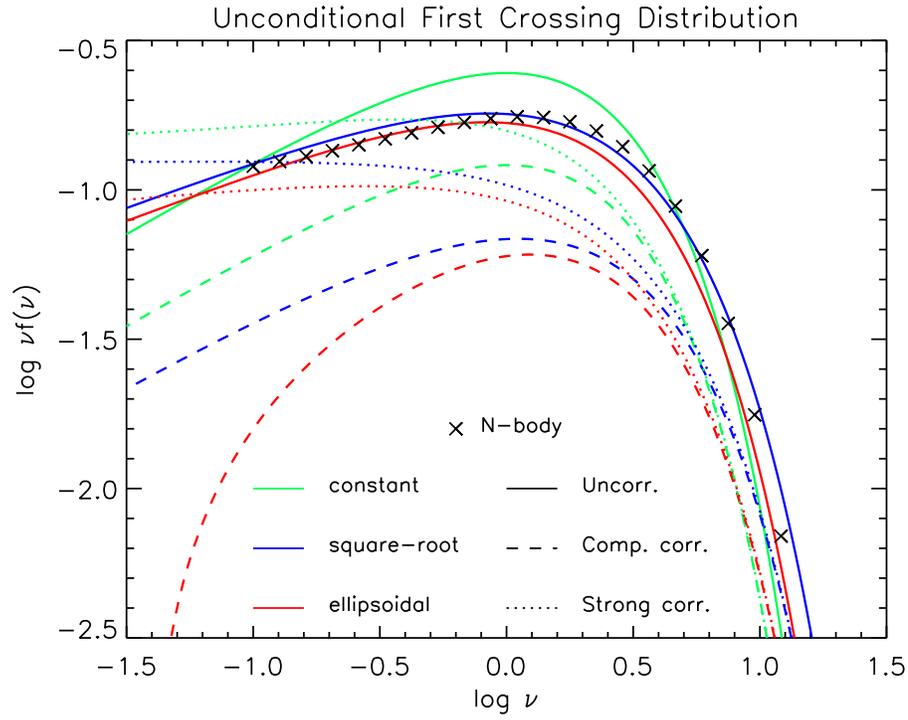}\caption{Comparisons between the unconditional
first crossing distribution with uncorrelated (solid), completely correlated
(dashed) and strongly correlated (dotted lines) steps, for the constant
(green), square-root (blue), and ellipsoidal (red) barriers. Crosses show the
outcomes of cosmological $N$-body simulations (fit to Tinker et al. 2008).}
\end{figure}

\clearpage
\begin{figure}
\epsscale{1.0}\plotone{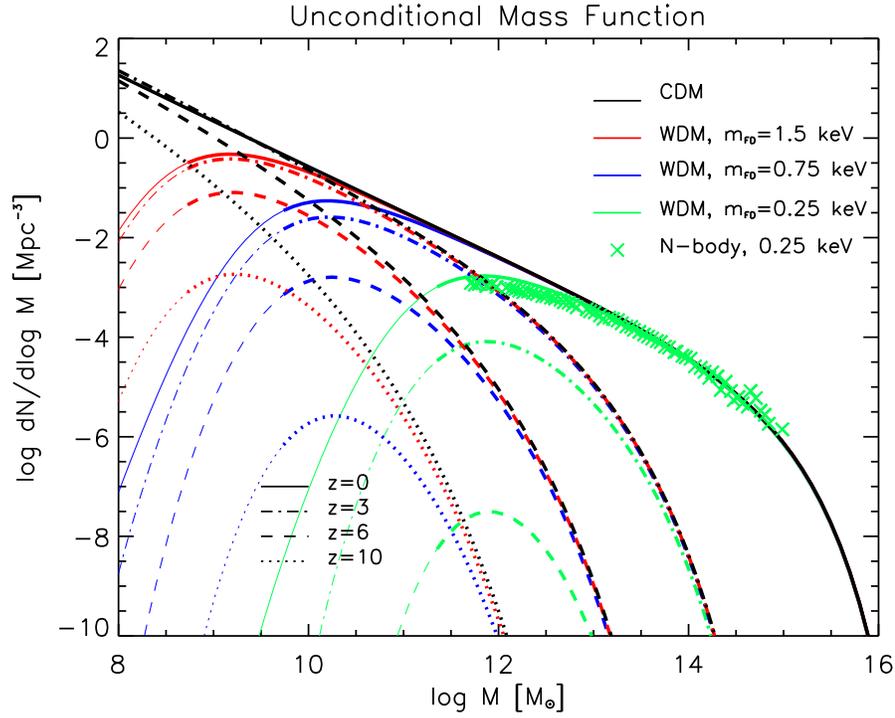}\caption{Unconditional mass function as a
function of halo mass at different redshifts, for the ellipsoidal barrier.
Black line refer to a standard Cold DM power spectrum, red line to a Warm DM with
Fermi-dirac mass $m_{\rm FD}=1.5$ keV, blue line to $m_{\rm FD}=0.75$ keV,
and green line to $m_{\rm FD}=0.25$ keV. The latter is compared at $z=0$ with
the outcome of the $N$-body simulation by Schneider et al. (2012; green
crosses) for the same mass $0.25$ keV. Lines are thick above, and thin below,
the free-streaming mass of the Warm DM particle.}
\end{figure}

\clearpage
\begin{deluxetable}{lcccccccc}
\tabletypesize{}\tablecaption{Main Results} \tablewidth{0pt}
\tablehead{\colhead{Results} && \colhead{Sections} && \colhead{Equations} &&
\colhead{Figures}}\startdata
Unconditional first crossing distribution$\dag$ && 3 && 20,21 && 3\\
Unconditional halo mass function$^\ddag$ && 2,3 && 9,20 && 4\\
Conditional first crossing distribution && 4 && 28 && 5\\
Conditional halo mass function && 4 && 28,29 && 6\\
Halo formation rates && 5.1 && 32,34 && 7\\
Halo creation rates && 5.1 && 35,38 && 8,10\\
Halo creation time distribution && 5.1 && 39,40 && 9\\
Halo growth history && 5.2 && 43 && 11,12\\
Halo bias && 5.3 && 44,45 && 13\\
\enddata \tablecomments{$\dag$ Results for correlated steps are
discussed in Appendix B and illustrated in Fig.~14. $^\ddag$ Results for Warm
DM power spectra are discussed in Appendix C and illustrated in Fig.~15.}
\end{deluxetable}

\end{document}